\documentclass[useAMS,usenatbib]{mn2e}

\usepackage{amsmath} 
\usepackage{subfigure} 
\usepackage{epsfig}
\usepackage{appendix}
\usepackage{xspace}
\usepackage{multirow}
\usepackage[T1]{fontenc}
\usepackage{url}
\usepackage{xcolor}
\usepackage{footnote}

\def\reff@jnl#1{{\rm#1\/}}
\def\aj{\reff@jnl{AJ}}         
\def\araa{\reff@jnl{ARA\&A}}      
\def\apj{\reff@jnl{ApJ}}        
\def\apjl{\reff@jnl{ApJ}}        
\def\apjs{\reff@jnl{ApJS}}       
\def\aap{\reff@jnl{A\&A}}        
\def\aapr{\reff@jnl{A\&A~Rev.}}     
\def\aaps{\reff@jnl{A\&AS}}       
\def\mnras{\reff@jnl{MNRAS}}      
\def\physrep{\reff@jnl{Physics Reports}}
\def\prd{\reff@jnl{Phys.Rev.D}}     
\def\prl{\reff@jnl{Phys.Rev.Lett}}   
\def\pasp{\reff@jnl{PASP}}       
\def\pasj{\reff@jnl{PASJ}}       
\def\nat{\reff@jnl{Nature}}       

\def\Sref#1{Section~\ref{#1}\xspace}
\def\srefa#1{Sections~\ref{#1}\xspace}
\def\srefb#1{\ref{#1}\xspace}
\def\Fref#1{Figure~\ref{#1}\xspace}
\def\frefa#1{Figures~\ref{#1}\xspace}
\def\frefb#1{\ref{#1}\xspace}
\def\Tref#1{Table~\ref{#1}\xspace}
\def\trefa#1{Tables~\ref{#1}\xspace}
\def\trefb#1{\ref{#1}\xspace}
\def\Eref#1{Equation~\ref{#1}\xspace}
\def\erefa#1{Equations~\ref{#1}\xspace}
\def\erefb#1{\ref{#1}\xspace}
\def\Aref#1{Appendix~\ref{#1}\xspace}
\def\eg{{\it e.g.\ }}
\def\ie{{\it i.e.\ }}
 
\def\phosim{{\sc PhoSim}\xspace}
\def\Phosim{{\sc \textbf{PHOSIM}}\xspace}
\def\opsim{{\sc OpSim}\xspace}
\def\imcat{{\sc imcat}\xspace}

\def\Neff{N_{\rm eff}}

\newcommand{\chihway}[1]{\textcolor{black}{#1}}

\def\kipac{KIPAC, Stanford University, 452 Lomita Mall, 
Stanford, CA 94309, USA}

\def\oxford{Department of Physics, University of Oxford, 
 Keble Road, Oxford, OX1 3RH, UK}
\def\ssl{Space Sciences Laboratory, University of California, 
 Berkeley, CA 94720, USA}
\def\purdue{Department of Physics, Purdue University, 
 West Lafayette, IN 47907, USA}
\def\uw{Department of Astronomy, University of Washington, 
Seattle, WA 98195}
\def\davis{Department of Physics, University of California, Davis, 
One Shields Avenue, Davis, CA 95616, USA}

\title[Spurious Shear in Weak Lensing with LSST]
{Spurious Shear in Weak Lensing with LSST}

\author[C.~Chang et al.]
{C.~Chang,$^{1}$\thanks{E-mail: chihway@slac.stanford.edu}
 S.~M.~Kahn,$^{1}$
 J.~G.~Jernigan,$^{2}$
 J.~R.~Peterson,$^{3}$
 \newauthor
Y.~AlSayyad,$^{4}$ 
Z.~Ahmad,$^{3}$
J.~Bankert,$^{3}$ 
D.~Bard,$^{1}$
A.~Connolly,$^{4}$ 
R.~R.~Gibson,$^{4}$ 
 \newauthor
K.~Gilmore,$^{1}$
E.~Grace,$^{3}$ 
M.~Hannel,$^{3}$
M.~A.~Hodge,$^{3}$ 
M.~J.~Jee,$^{6}$
L.~Jones,$^{4}$ 
 \newauthor
S.~Krughoff,$^{4}$ 
S.~Lorenz,$^{3}$
P.~J.~Marshall,$^{5}$ 
S.~Marshall,$^{1}$ 
A.~Meert,$^{3}$
S.~Nagarajan,$^{3}$ 
 \newauthor
E.~Peng,$^{3}$ 
A.~P.~Rasmussen,$^{1}$
M.~Shmakova,$^{1}$ 
N.~Sylvestre,$^{3}$
N.~Todd,$^{3}$ 
M.~Young$^{3}$ \\ \\
$^{1}$\kipac\\
$^{2}$\ssl\\
$^{3}$\purdue \\
$^{4}$\uw\\
$^{5}$\oxford\\
$^{6}$\davis}

\begin{document}
\date{Accepted 2012 October 16.  Received 2012 August 28; in original form 2012 June 6 }

\pagerange{\pageref{firstpage}--\pageref{lastpage}} \pubyear{2011}

\maketitle

\label{firstpage}

\begin{abstract}

The complete 10-year survey from the Large Synoptic Survey Telescope (LSST) 
will image $\sim$ 20,000 square degrees of sky in six filter bands every few nights, bringing 
the final survey depth to $r\sim27.5$, with over 4 billion well measured galaxies. To take full 
advantage of this unprecedented statistical power, the systematic errors associated with weak 
lensing measurements need to be controlled to a level similar to the statistical errors. 

This work is the first attempt to quantitatively estimate the absolute level and statistical 
properties of the systematic errors on weak lensing shear measurements due to the 
most important physical effects in the LSST system via high fidelity ray-tracing 
simulations. 
\chihway{
We identify and isolate the different sources of algorithm-independent, 
\textit{additive} systematic errors on shear measurements for LSST and predict their impact 
on the final cosmic shear measurements using conventional weak lensing analysis 
techniques. We find that the main source of the errors comes from an inability to adequately 
characterise the atmospheric point spread function (PSF) due to its high frequency spatial 
variation on angular scales smaller than $\sim10'$ in the single short exposures, which 
propagates into a spurious shear correlation function at the $10^{-4}$--$10^{-3}$ level on 
these scales. With the large multi-epoch dataset that will be acquired by 
LSST, the stochastic errors average out, bringing the final spurious shear correlation function 
to a level very close to the statistical errors. Our results imply that the cosmological constraints 
from LSST will not be severely limited by these algorithm-independent, additive systematic 
effects. }

\end{abstract}



\section{Introduction}
\label{sec:Introduction}

Weak gravitational lensing, or weak lensing for short, is one of the most powerful 
tools for probing dark matter and dark energy \citep{2006APS..APR.G1002A}. 
Distorted by intervening large-scale structures, the otherwise randomly oriented 
galaxy images encode signatures of dark matter and dark energy in a statistical 
way, namely through cosmic shear. For a review of weak lensing, see, for example, 
\citet[][hereafter BS01]{2001PhR...340..291B}. The lensing power spectrum 
provides a unique tool to distinguish between different cosmological models 
\citep{1997ApJ...484..560J, 1998ApJ...498...26K, 1999ApJ...514L..65H}.

Since the first detections of the cosmic shear signal by several independent 
groups \citep{2000Natur.405..143W, 2000MNRAS.318..625B, 
2000astro.ph..3338K}, there has been an explosion of research activity in this 
field. The most recent analyses have shown that state-of-the-art weak 
lensing surveys are already probing interesting regions of the dark energy 
parameter space \citep{2006A&A...452...51S, 2007MNRAS.381..702B, 
2007A&A...468..859H, 2010A&A...516A..63S, 2011arXiv1111.6622L, 
2011arXiv1112.3143H}. However, a major limitation of these existing surveys has 
been their relatively small sky coverage, which results in an insufficient number of 
galaxies to average out their random shapes and orientations (\ie to reduce the 
so-called ``shape noise''). Cosmic shear measurements to date are limited by such 
statistical errors. 

As a result, several projects are attempting to overcome this fundamental limitation 
by significantly increasing the sky coverage. 
\chihway{The Dark Energy Survey\footnote{\url{http://www.darkenergysurvey.org/}}, 
the Kilo Degree Survey\footnote{\url{http://kids.strw.leidenuniv.nl/}}, }
Hyper Suprime Cam\footnote{\url{http://www.astro.princeton.edu/~rhl/HSC/}}, 
LSST\footnote{\url{http://www.lsst.org/}}\citep{2002SPIE.4836...10T} and 
Euclid\footnote{\url{http://sci.esa.int/science-e/www/area/index.cfm?fareaid=102}} 
projects have all been explicitly designed for weak lensing investigations. The primary 
improvement of these projects over previous ones is that the cameras they 
incorporate have very large fields of view, which leads to a much larger survey 
area and a dramatic improvement in the statistical power of the dataset 
\citep{2007MNRAS.381.1018A}. When statistical errors become negligibly 
small in these future surveys, systematics errors become a primary concern. 

For weak lensing, there are systematic errors associated with physical effects 
in the atmosphere and the telescope, and with imperfect algorithms used in the
analysis. In this paper, we are mostly interested in quantitatively characterising 
the former. Systematic errors due to imperfect algorithms are in principle 
reducible, and will certainly shrink as we gain experience with the near-term upcoming 
surveys. However, physical effects that are inherent to the system and independent of 
specific weak lensing algorithms are irreducible and most likely will determine the 
ultimate limits on cosmological constraints derived from weak lensing. 

In the past, the effects of different sources of systematic errors on cosmic shear 
measurements have usually been calculated by assuming some hypothetical power 
spectrum for the spurious shear, often in simple functional forms for analytical 
calculations \citep[][hereafter AR08]{2008MNRAS.391..228A}. However, these functional 
forms may not be well motivated by physics. We make the first attempt to approach the 
problems in a bottom-up way and \textit{simulate} the actual measurements to predict the 
level of systematic errors from first principles. We use LSST as our benchmark survey in 
this work, but many of the results are general or scalable to other future weak lensing 
surveys. The LSST Photon Simulator, or \phosim \citep[][]{P10,P12, 2010SPIE.7738E..53C} 
is used in this work to generate realistic LSST images for this study. In this way, we are able 
to measure quantitatively the systematic errors generated from various physical effects in a 
controlled way. 

\chihway{
Note that in this paper we only discuss the case of additive shear systematics 
associated with the projected two-point correlation function on a limited range of angular 
scales (within the field of a single focal plane). We do not consider weak lensing 
tomography \citep{1999ApJ...522L..21H} or higher order statistics 
\citep{2003A&A...397..809S, 2005A&A...431....9S}. The use of these other statistics can 
impose additional requirements on the level of systematic errors, but on the other hand, 
if the information is combined properly, it also has the potential of mitigating particular 
systematic errors that are only present in the projected two-point correlation function.}


The paper is organised as follows. A brief review of the canonical framework of 
weak lensing is  given in \Sref{sec:Theory_WL}. In \Sref{sec:ImSim} we present a 
short  introduction to LSST and our simulation tool. In \Sref{sec:Framework_EllipSys}, 
we lay out a framework for classifying the different physical effects that induce errors 
in shape measurements. In \Sref{sec:EllipSys}, the different sources of errors and their 
correlation properties are quantified using simulations. Possible sources of spurious 
shear signals after correcting for the PSF effects are discussed in 
\Sref{sec:Framework_ShearSys}, while the results from simulations are presented in 
\Sref{sec:ShearSys}. In \Sref{sec:Discussion}, we discuss the prospect of combining 
multiple exposures, the implications for the determination of cosmological constraints, 
and the effect of some of our assumptions on our results. We conclude in 
\Sref{sec:Conclusion}.


\section{Weak lensing notation and measurements}
\label{sec:Theory_WL}

In the presence of weak lensing, a galaxy image, having some intrinsic shape, is 
first sheared by the gravitational potential along the line-of-sight, then convolved with 
the atmospheric and instrumental PSF before being measured. As an observer, we 
want to reverse this process: measure the shape of a galaxy from a noisy image, 
correct for the PSF effects to infer the shear through an estimator, and finally calculate 
different statistics that are sensitive to cosmology using the shear estimator. For details 
on the weak lensing formalism, as well as predictions of weak lensing signals from 
different cosmological models, see BS01.

For this work, following the steps in a data reduction process, we ask the 
following questions: (1) How do the different physical effects change the measured galaxy 
shape before any PSF correction has been made? (\Sref{sec:Framework_EllipSys}, 
\Sref{sec:EllipSys}) (2) To what level can these PSF effects be corrected to infer the correct 
shear using a conventional algorithm? (\Sref{sec:Framework_ShearSys}, \Sref{sec:ShearSys}) 
(3) With only the information from two-point shear correlation functions, how do the effects in 
(1) and (2) scale in the final combined dataset and what does that imply in terms of uncertainties 
in our predicted cosmological model? (\Sref{sec:Discussion})

\subsection{Weak lensing notation}
\label{sec:Theory_WL_notation}
Throughout the paper we use the following definition for the complex ``ellipticity spinor'',  
$\boldsymbol{\varepsilon}=\varepsilon_{1}+i\varepsilon_{2}$, to parametrise the shapes of 
objects:
\begin{equation}
  \varepsilon_{1} =\frac{I_{11}-I_{22}}{I_{11}+I_{22}} \; , 
  \; \varepsilon_{2} =\frac{2I_{12}}{I_{11}+I_{22}} \;.
  \label{eq:ellipticity}
\end{equation}

\noindent where $I_{ij}$ are normalised moments of the object's light intensity profile 
$f(x_{1},x_{2})$, weighted by a Gaussian filter $W(x_{1},x_{2})$ to reduce noise: 
\begin{equation} 
  I_{ij}=\frac{\int \int dx_{1}dx_{2}W(x_{1},x_{2})f(x_{1},x_{2})x_{i}x_{j}}
  {\int \int dx_{1}dx_{2}W(x_{1},x_{2})f(x_{1},x_{2})}, \; i,j=1,2  \;.
\label{eq:moments}
\end{equation}

\noindent where the width of $W(x_{1},x_{2})$ is chosen to give the maximum 
signal-to-noise ratio for each individual object. 

In this paper, if not otherwise specified, boldface symbols indicate the complex quantities and 
the magnitude of the complex quantity is specified using the corresponding regular-font symbol 
(\eg $ \varepsilon=|\boldsymbol{\varepsilon}|=\sqrt{\varepsilon_{1}^{2}+\varepsilon_{2}^{2}} $).
A similar notation is used to parametrise shear $\boldsymbol{\gamma}$, where we have 
$\boldsymbol{\gamma}=\gamma_{1}+i\gamma_{2}$. 

We also adopt the standard definitions for calculating the correlation function 
$ \xi_{+,\boldsymbol{XX}}(\theta)$ and power spectrum $C_{\boldsymbol{X}}(\ell)$:
\begin{equation}
  \xi_{+,\boldsymbol{XX}}(\theta) = \langle X_{t}(\theta_{0}) 
  X_{t}(\theta_{0}+\theta) \rangle + \langle 
  X_{\times}(\theta_{0}) X_{\times} (\theta_{0}+\theta ) \rangle \;,
  \label{eq:cf}
\end{equation}
\begin{equation}
  C_{\boldsymbol{X}}(\ell)=2 \pi \int^{\infty}_{0} d\theta \theta \xi_{+,\boldsymbol{XX}}(\theta)
  J_{0}(\ell\theta) \; ,
  \label{eq:cf2ps}
\end{equation}
\begin{equation}
  \xi_{+,\boldsymbol{XX}}(\theta)=\frac{1}{2\pi}  \int^{\infty}_{0} d \ell \ell C_{\boldsymbol{X}}(\ell)J_{0}(\ell\theta) \; ,
  \label{eq:ps2cf}
\end{equation}

\noindent where $\boldsymbol{X}$ is a complex spinor (\eg ellipticity 
$\boldsymbol{\varepsilon}$ or shear $\boldsymbol{\gamma}$) and the subscripts $t, \times$ 
indicate an isotropised decomposition of $\boldsymbol{X}$ along the line connecting a 
particular pair of galaxies. If $\boldsymbol{X}$ is measured in an arbitrary Cartesian coordinate 
system $\boldsymbol{X}=X_{1}+iX_{2}$ with 1,2 denoting the two axes, then the rotated shear 
is calculated via $X_{t}=-Re(\boldsymbol{X} e^{-2i\varphi})$ and 
$X_{\times}=-Im(\boldsymbol{X} e^{-2i\varphi})$, where $\varphi$ is the argument of the vector 
connecting the pair of galaxies. The angle brackets $\langle \rangle$ indicate an average over 
all galaxy pairs separated by $\theta$ (with one galaxy located at some $\theta_{0}$). $J_{0}$ 
is the zeroth-order Bessel function of the first kind. We will use $\xi_{XX}$ as shorthand for 
$\xi_{+,\boldsymbol{XX}}(\theta)$ for the rest of the paper. For the simulations in this work, we 
look at angular scales up to the scale of the full LSST focal plane ($\sim3$ degrees). 

\subsection{Analysis tools}
\label{sec:Theory_WL_measurement}
In all of our analyses of the simulated images, we use the software package {\sc Source Extractor} 
\citep{1996A&AS..117..393B} for object detection. We set the following configuration 
parameters: $\rm DETECT\_MINAREA=5$ and $\rm DETECT\_THRESH=1.5$. 

\chihway{Background estimation, shape measurement, PSF correction and shear estimation 
were done through the software package 
{\sc IMCAT}\footnote{\url{http://www.ifa.hawaii.edu/~kaiser/imcat/download.html}} based on the 
algorithm derived in \citet{1995ApJ...449..460K}, \citet{1997ApJ...475...20L} and 
\citet{1998ApJ...504..636H}, commonly known as KSB. The {\sc IMCAT} parameters e[0], e[1], 
gamma[0] and gamma[1] correspond to the ellipticity and shear components $\varepsilon_{1}$, 
$\varepsilon_{2}$, $\gamma_{1}$ and $\gamma_{2}$ respectively, while the {\sc IMCAT} 
parameter $\rm r_{g}$ is used for the width of $W(x_{1},x_{2})$ in \Eref{eq:moments}. Our 
specific implementation of KSB is similar to the ``ES2'' method in \citet{2007MNRAS.376...13M}. 
We describe briefly the KSB formulae in \Aref{sec:KSB}.}


\section{LSST and \Phosim}
\label{sec:ImSim}

The LSST survey will be the most powerful ground-based weak lensing survey planned for the 
coming decade. Its revolutionary scale will likely lead the next generation of optical survey 
designs. We therefore believe that using LSST as the target for this study will enable us to 
capture the most important issues for future weak lensing surveys. 

\subsection{LSST design parameters}
\label{sec:ImSim_LSST}
\begin{figure}
  \begin{center}
  \includegraphics[height=3in]{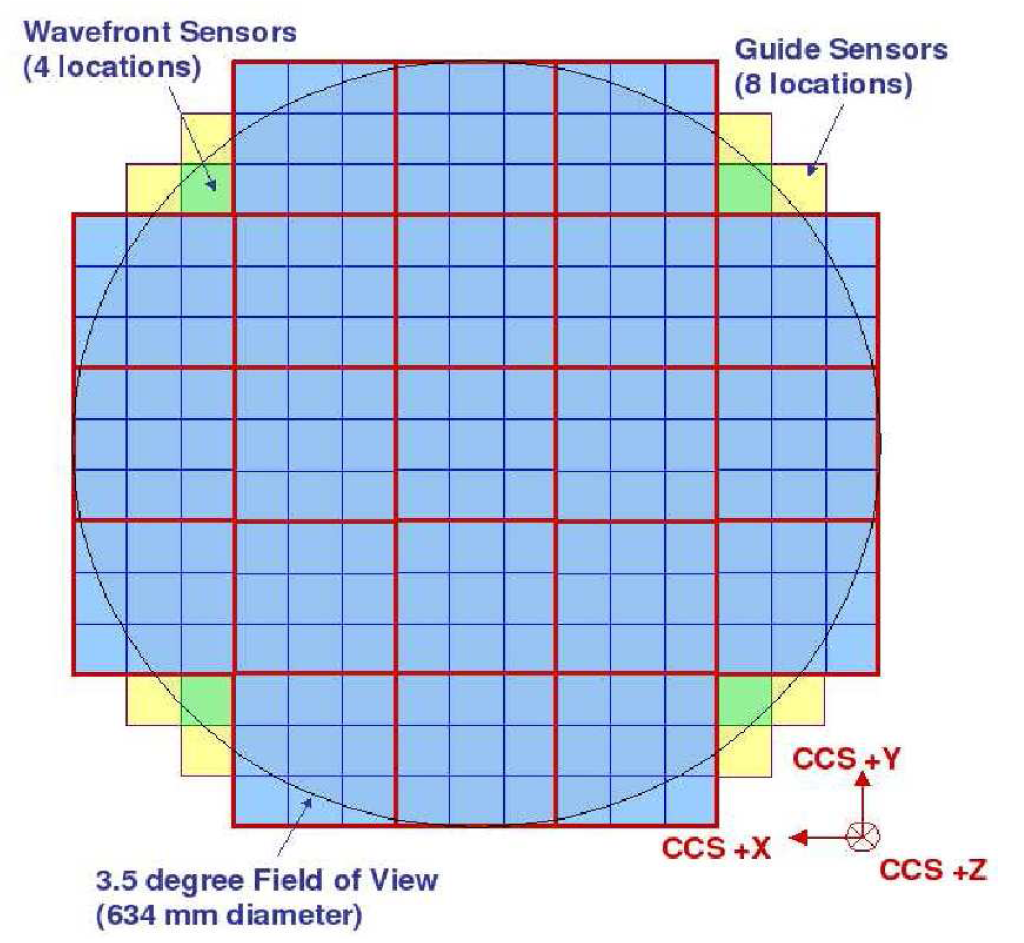}
  \end{center}
   \caption{Layout of the LSST focal plane taken from \citet{2008arXiv0805.2366I}. The red and 
   blue thin lines indicate the boundaries for each sensor, where the blue area is the 189 science 
   sensors we use in our focal-plane size simulations. The ellipticity maps from our simulations 
   shown in \frefa{fig:ns_maps} and \frefb{fig:s_maps} correspond to ellipticity values in these 
   areas.}
   \label{fig:focalplane}
\end{figure}

The optical design of LSST is optimised to cover as much sky as possible while maintaining 
good image quality \citep{2008arXiv0805.2366I}. The 8.4-meter aperture and the 
9.5-degree$^{2}$ field of view combine to an \'etendue of $\sim$319.5 m$^{2}$degree$^{2}$, 
which is over 10 times larger than that of any previous survey facility. The heart of the 
instrument is a 64-cm-diameter, 3.2-giga-pixel focal plane. The focal plane is tiled with 189 
CCD sensors, each with 4k$\times$4k, 10 $\rm \mu m$ square pixels (each pixel corresponds 
to an angular scale of 0.2"). The layout of the focal plane geometry is shown in \Fref{fig:focalplane}.

Good image quality is one of the key components to weak lensing measurements. To ensure 
that over the entire survey period the instrumental effects that degrade the image quality are 
kept under control, LSST incorporates an Active Optics System (hereafter AOS), which adjusts 
the figures and positions of the three reflective optics and the orientation of the camera to 
correct the wavefront errors. 

LSST will take images approximately every 15 -- 20 seconds, covering the entire available 
hemisphere every few days in six optical filter bands. 
\chihway{
Each of the 2 consecutive 15-second exposures (separated by 4-second readout and shutter 
open/close) is called a \textit{visit}. The 2 exposures in 
a visit will be taken on the same field. From visit to visit, the telescopes will then point to different 
fields in order to achieve the very wide sky coverage.} The 10-year survey will generate 
an unprecedentedly large amount of data (nearly two thousand 15-second exposures on each 
field across the 20,000 degree$^{2}$ sky). For cosmic shear measurements, this means 
reducing the statistical errors from shape noise and cosmic variance by orders of magnitude. 
As a result, understanding the sources of systematic errors in these data will undoubtedly be 
a major challenge for LSST.

\subsection{LSST observation parameters}
\label{sec:ImSim_opsim}
We extract from catalogues generated by the LSST Operations Simulator \citep[][hereafter \opsim]
{2010AAS...21540105K} information about the the observing conditions in an expected LSST 
weak lensing dataset. \opsim simulates the atmosphere and night sky conditions for individual 
exposures at the LSST site over 10 years based on weather models, telescope models and 
optimisation of the survey strategy. 

From previous work, it is known that only images with the best image quality 
contribute to the cosmic shear signal \citep{2006ApJ...647..116H}. As a result, to estimate more 
accurately the ``typical'' observing parameters for images that contribute to the final cosmic shear 
measurement for LSST, we take mediums of the major observation parameters in a subset of 
the full \opsim catalogue. This subset consists of the 50\% of the $r$-band (552--691 nm) data that 
give the best image quality. 
From this process we define the ``fiducial observing parameters'' for a 10-year LSST weak 
lensing dataset as listed in \Tref{table:fiducial_params}. Note that the parameters in 
\Tref{table:fiducial_params} are specified (when applicable) for $r$ band, while for real 
weak lensing analyses, $i$-band (691--818 nm) images are likely to be used as well. We 
carry out all the analyses presented here in $r$ band, but extrapolate the results to $i$ 
band (see \Sref{sec:Combine}), knowing that the image quality and observing parameters 
are similar in both bands. In \Tref{table:fiducial_params}, the parameter $a_{\rm opt}$ 
is the designed instrumental PSF size specified in 
\cite{SRD}\footnote{\url{http://www.lsst.org/files/docs/SRD.pdf}} at the elevation that 
corresponds to the median airmass. 

\begin{table}
  \centering
  \caption{Fiducial observing parameters. All parameters are specified for the LSST $r$ band 
  (552-691 nm). These observing parameters are chosen to be typical for obtaining images for 
  weak lensing measurements. The numbers are calculated 
  from the medium values of the best 50\% $r$-band exposures. Note that $a_{\rm atm}$ 
  includes the airmass contribution.} 
  \label{table:fiducial_params}  
  \begin{tabular}{c c c}
  \hline
  Parameter name    &   Description                  &  Fiducial value \\ \hline \hline
      $a_{\rm atm}$     &  atmospheric seeing   &  0.56"                  \\ \hline
      \multirow{2}{*}{$a_{\rm opt}$}     &  instrumental  &   \multirow{2}{*}{0.42"}    \\ 
           &  PSF FWHM &                 \\ \hline
      $X_{\rm air}$      &  airmass                         &  1.2                    \\ \hline
      $B_{\rm sky}$     &  sky background          &  640 counts/pixel      \\ \hline
      $t_{\rm exp}$      &  exposure time             &  15 seconds              \\ \hline
      $b$                       &  Galactic latitude          &  $-60^{\circ}$     \\ \hline
      $N$                      &  number of exposures &  184     \\ \hline
  \end{tabular}
\end{table}

\subsection{\Phosim}
\label{sec:ImSim_imsim}
To study in detail the expected systematic errors in weak lensing measurements for this 
survey, existing data from other projects are insufficient in both the data quality and 
quantity. As a result, simulations become the only way to investigate such problems before 
the telescope is built. For this study in particular, simulations also enable us to trace the 
individual sources of systematic errors in a controlled and bottom-up fashion, which is 
almost impossible to achieve with real data. A few unique features of the simulation 
process in this work should be emphasised: First, all the physical effects that introduce 
systematic errors in the shape measurements can be separately 
turned ``on'' and ``off''. Therefore, we have full control over which actual physical effects are 
dominant in determining the image shape error. Second, the PSF at the location of 
a galaxy image can be known \textit{exactly} by simulating an image with a bright source 
at that same location. Finally, all physical processes in these numerical experiments are 
\textit{reproducible}. 

In an earlier attempt to simulate LSST as a complete system using a modified version of 
existing optics software \citep{2011PASP..123..596J}, the potential power of studying these 
issues via simulations has been demonstrated. In this work, we take the analysis one step 
further and invoke \phosim as our primary tool for generating simulated images. Unlike 
the software used in \citet{2011PASP..123..596J}, \phosim is a 
set of custom-made software designed specifically to represent the LSST's performance, and 
simultaneously incorporate many aspects of the project design (\eg data management software 
development and scientific studies). \phosim adopts a photon-by-photon Monte Carlo fast 
ray-tracing algorithm, which generates images expected for LSST with very high fidelity. In 
collaboration with the LSST instrumentation teams and multiple science groups, the \phosim 
software has been continuously updated and cross-checked to track the most current hardware 
developments. 

\phosim is part of the end-to-end LSST Image 
Simulator\footnote{\url{http://lsst.astro.washington.edu/}} ({\sc ImSim}). {\sc ImSim} simulates 
the forward process from cosmological models to realistic astronomical images, in which 
\phosim is responsible for the last part in this process -- the photon propagation from top of the 
atmosphere down to the CCD sensors and the signal readout. {\sc ImSim} begins with a catalogue 
of celestial sources based on large cosmological N-body simulations and detailed Milky Way and 
Solar System models. A realistic observing environment is then set up by using parameters 
predicted by the \opsim catalogue. \phosim then simulates the final ``exposure'' by tracing individual 
photons from objects in the catalogue for that part of the sky, through the atmosphere, the telescope, 
and into the camera to form an image that retains all the major characteristics we anticipate in the 
LSST data. 

We use \phosim version 3.0 to do all the analyses in this paper. To provide sufficient background 
material to interpret the results, we describe briefly in \Aref{sec:ImSim_model_app} the major 
physical models in \phosim, and refer to \citet{P12} for further details.


\section{Sources of ellipticity errors}
\label{sec:Framework_EllipSys}

Assume the PSF has some finite size $a$ and we measure a PSF-convolved galaxy to have 
ellipticity $\boldsymbol{\varepsilon}^{m}$, then this $\boldsymbol{\varepsilon}^{m}$ can be 
broken down to an intrinsic component, a shear component and an additional component from 
various physical effects associated with counting statistics, the atmosphere and the 
telescope/camera system. If we assume that all these three components are evaluated 
for \textit{the same measured galaxy size} and we are only interested in the small changes in the 
\textit{anisotropy} of the galaxy shape\footnote{Because the measured ellipticity of a galaxy after 
convolution with a PSF depends nonlinearly on the width of the PSF, we make this assumption 
and only investigate the \textit{anisotropic} part of the ellipticity change, which can be viewed as 
linear.}, we can write out the following relation in linear additive terms: 
\begin{equation}
  \boldsymbol{\varepsilon}^{m}=\boldsymbol{\varepsilon}^{i}
  +\frac{2}{\alpha}\boldsymbol{\gamma}+\boldsymbol{\varepsilon}^{s}\; ,
  \label{eq:measured_e}
\end{equation}
\noindent where the first term is the ellipticity of the galaxy convolved with a circular PSF, the second 
term is the change in $\boldsymbol{\varepsilon}^{m}$ due to shear and the last term is the change in 
$\boldsymbol{\varepsilon}^{m}$ due to other physical effects. $\alpha$ is a scaling factor to first 
``de-weight'' the ellipticity calculated from the weighted moments (\Eref{eq:moments}) and then 
account for the effect of the finite-size PSF. The factor of 2 in the second term converts shear into 
ellipticity (BS01). For infinite resolution (PSF $\sim$ delta function) and ellipticities calculated from 
unweighted moments ($W=1$), we have $\alpha=1$. For infinite resolution and ellipticities calculated 
from weighted moments, $\alpha$ is equivalent to two times the shear responsivity 
$P_{\alpha\beta}^{\gamma}$ defined in \cite{1998ApJ...504..636H}.

The correlation function for the measured ellipticity can thus be written out as:
\begin{align}
  \xi_{\varepsilon^{m}\varepsilon^{m}}=&\xi_{\varepsilon^{i}\varepsilon^{i}}
  +\frac{4}{\alpha^{2}}\xi_{\gamma\gamma}+\xi_{\varepsilon^{s}\varepsilon^{s}} \notag \\
  &+2(\frac{2}{\alpha}\xi_{\varepsilon^{i}\gamma}+\frac{2}{\alpha}\xi_{\gamma\varepsilon^{s}}+
  \xi_{\varepsilon^{i}\varepsilon^{s}}) \; .
  \label{eq:measured_Cee}
\end{align}

\subsection{Non-stochastic and stochastic errors in ellipticity measurements}
\label{sec:s_ns}
Here we present a concept for classifying $\boldsymbol{\varepsilon}^{s}$ similar to that in 
\citet{2006JCAP...02..001J}. This classification scheme is especially important for analyses of 
multi-epoch datasets such as LSST -- this is the first step towards understanding the nature of 
different sources of systematic errors in shear measurements. Two major classes of physical 
effects combine to give $\boldsymbol{\varepsilon}^{s}$. We use the terms ``non-stochastic'' 
and ``stochastic'' 
to refer to these two classes of errors.

Non-stochastic errors are those that are either fixed in space and time, or vary with characteristic 
patterns over multiple exposures. Stochastic effects, on the other hand, induce errors that change 
randomly from exposure to exposure with no correlation in time. For non-stochastic errors, 
because they show repeated patterns from frame to frame, one does not benefit from averaging 
over multiple independent exposures; however, this repeating feature also means that they potentially 
can be characterised very well when one properly combines the multi-epoch dataset. Stochastic errors 
are exactly the opposite: randomness implies one can only model them with data from limited 
information in a single exposure, but via some form of averaging of the multiple exposures on the same 
field, the errors are likely to cancel each other. 

\begin{savenotes}
\begin{table}
\begin{center}
    \caption{Classification of the major physical effects that introduce errors in shape measurements. 
    The difference between non-stochastic and stochastic optics errors is described in 
    \Sref{sec:Framework_EllipSys}. }
    \begin{tabular}{c c}
    \hline
     Non-stochastic effects & Stochastic effects\\ \hline \hline
     Optics design & Counting statistics\\ 
     Non-stochastic optics errors & Stochastic optics errors  \\ 
      & Tracking errors \\
      & Atmospheric effects\footnote{\chihway{Atmospheric effects here does not include effects that 
      change the measured size of the galaxy such as variation in the seeing and airmass. Instead we 
      use the median seeing and airmass as listed in \Tref{table:fiducial_params} through the paper.}}      \\ \hline 
    \end{tabular}
    \label{table:classification}
\end{center}
\end{table}
\end{savenotes}

In \Tref{table:classification}, we identify the major non-stochastic and stochastic effects that are 
modeled in \phosim and are most likely to contribute to $\varepsilon^{s}$. \chihway{We also provide in 
\Aref{sec:ImSim_model_app} brief descriptions of how each of these effects is modelled in \phosim.} 
There are some physical effects that may be present and are not yet modeled in the current \phosim, but 
we believe they will not contribute significant.ÊThe effects listed in \Tref{table:classification} should 
comprise the great majority of the sources of error for weak lensing measurements with LSST.
  
In most existing weak lensing data, non-stochastic effects dominate the error; therefore 
the origins of these systematic errors are historically better understood. For example, 
\citet{2008arXiv0810.0027J} were able to model the PSF patterns of telescope aberrations with low order 
functional forms. Stochastic effects, being relatively small in existing data, have not been studied in detail. 
Only a few pioneering studies have tried to understand the stochastic effects under simple cases: 
\citet{2008A&A...484...67P} and \citet{2010MNRAS.403..673Z} studied the noise contribution to shape 
measurement due to counting statistics and pixelation, while \citet{2007ApJ...662..744D} investigated the 
atmosphere-induced ellipticity and its time dependence.
 
\chihway{Note that this classification scheme is only valid under a well-defined survey since it depends on the 
cadence of the survey and other operational issues. In this paper we are assuming the as designed LSST 
survey mode (\Sref{sec:ImSim_LSST}), where the minimum time between consecutive visits on the same 
patch of sky is approximately 30 minutes \citep{SRD}. This means that the telescope has experienced at least 
$\sim$50 different pointings between the two consecutive visits and is looking through a very different column of 
atmosphere each time. Under this scenario, most of the physical effects we discussed are truly stochastic, 
or at least stochastic to a very high level between visits. For the two exposures in the same visit, due to the 
close separation in time, stochasticity is not guaranteed for all effects -- we discuss in \Aref{sec:Neff} how this 
factor may be estimated for the spurious shear correlation function in the combined dataset.}
  
\subsection{Practical considerations}
\label{sec:Framework_circular_gaussians}
    
Real galaxies have intrinsic shapes, and will be subject to cosmic shear, so that in \Eref{eq:measured_e}, 
$\boldsymbol{\varepsilon}^{i}$ and $\boldsymbol{\gamma}$ are not generally equal to zero. To average 
over these effects at the statistical level sampled by LSST, we would need to simulate $\sim$200 images 
of roughly four billion galaxies in each test. This is computationally impractical, so we adopted a simpler 
approach described below. 
    
Note from \erefa{eq:measured_e} and \erefb{eq:measured_Cee} that if we set up simulations so that 
$\boldsymbol{\varepsilon}^{i}=\boldsymbol{\gamma}=0$, we can avoid the contribution from 
$\boldsymbol{\varepsilon}^{i}$ and $\boldsymbol{\gamma}$ in the observable $\boldsymbol{\varepsilon}^{m}$, 
and directly measure $\boldsymbol{\varepsilon}^{s}$ and $\xi_{\varepsilon^{s}\varepsilon^{s}}$ unambiguously. 
This suggests that our problem is equivalent to asking the following question:\\

\textit{Under zero shear, what is the anisotropic component of the spurious ellipticity $\boldsymbol{\varepsilon}^{s}$ 
induced by a certain physical effect on a circular object of $\boldsymbol{\varepsilon}^{i}=0$ and what are the 
correlation properties $\xi_{\varepsilon^{s}\varepsilon^{s}}$ of those spurious ellipticities?}\\

That is, we do not measure the ellipticity on a fully realistic galaxy population with a distribution of shapes, sizes 
and brightnesses; instead, simple circular ``galaxies'' are used as ``test particles'' for the entire population of 
galaxies. We show below that this approach is justified for our purposes. 

\begin{figure}  
  \begin{center}
  \includegraphics[height=2.3in]{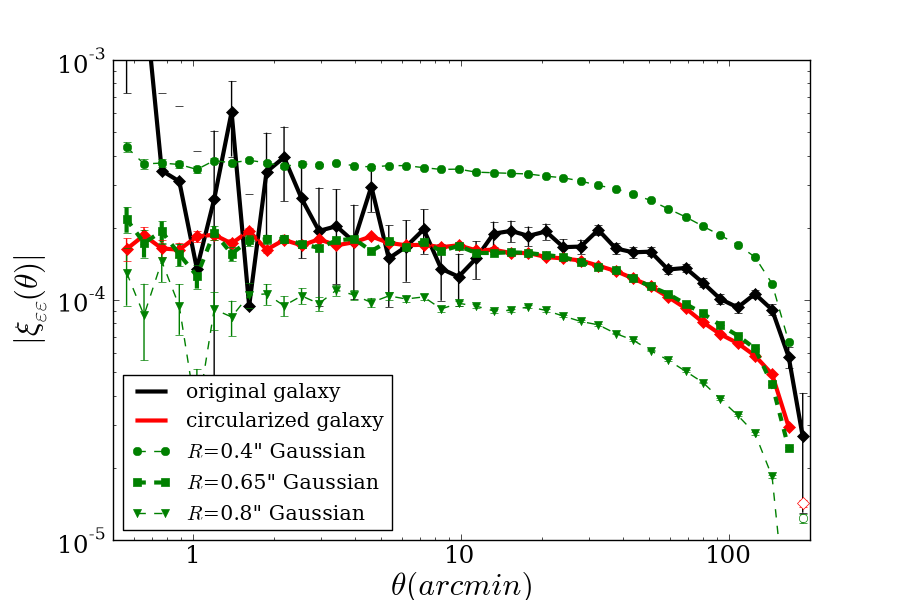} 
  \caption{Ellipticity correlation function for different samples of simulated galaxies. Black shows the 
  ellipticity correlation function for a realistic galaxy sample, while red shows the same function measured 
  from the circularised counterpart of the realistic galaxy sample. Green curves are measured from 
  samples of circular Gaussian shapes of different sizes. The green curve measured from circular Gaussians 
  of 23rd $r$-band magnitude and $R$=0.65" (green square) agree approximately in level and shape with 
  the black and red curves, where $R$ is the FWHM size of the Gaussian profile. This demonstrates (1) shape 
  noise is uncorrelated and (2) we can use the circular Gaussian shape with $r$-band magnitude 23 and 
  $R$ =0.65" as the ``fiducial galaxy'' to measure the spatial correlations of the response of the entire galaxy 
  population.}
  \label{fig:nulltest_cf}  
  \end{center}
\end{figure}
 
In \Fref{fig:nulltest_cf}, we simulate a representative galaxy sample from the \phosim sky catalogue and measure 
the ellipticity correlation function $\xi_{\varepsilon^{m}\varepsilon^{m}}$ from a typical single exposure. We then 
``circularise'' these galaxy images at the input catalogue level before entering the atmosphere 
so that they retain all the characteristics, such as size, brightness, redshift and spectral energy distribution 
(SED) in the original sample, but lose the shape information. Although the original sample shows a noisier 
ellipticity correlation function, the circularised sample roughly agrees with it in both level and shape. 
\chihway{
The agreement between the ellipticity correlation functions measured from the original galaxies and the 
circularised galaxies demonstrates that shape noise is not spatially correlated; thus it should play no role in the 
correlation function as expected. In other words, we have 
$\xi_{\varepsilon^{i}\varepsilon^{i}}=\xi_{\gamma\gamma}=\xi_{\varepsilon^{i}\gamma}=\xi_{\varepsilon^{s}\gamma}=0$ 
for both samples. The slightly lower red curve is mainly due to the small $\xi_{\varepsilon^{i}\varepsilon^{s}}$ term 
that is present only in the original galaxy sample. We show that we can isolate $\xi_{\varepsilon^{s}\varepsilon^{s}}$ in 
the $\xi_{\varepsilon^{m}\varepsilon^{m}}$ using the circularised galaxy sample. }
    
\begin{table}
\centering
  \caption{Fiducial galaxy characteristics specified in $r$ band. The representative sample of weak lensing 
  galaxies can be collapsed into the fiducial galaxy and reproduce the same ellipticity correlation function 
  -- on average, the fiducial galaxy reacts to the PSF effects the same way as the population of galaxies.} 
  \begin{tabular}{c c c}
  \hline
  Parameter name    &   Description               &  Fiducial value\\ \hline \hline
      $m$                      &   AB magnitude           & 23 \\ \hline
      $S$                       &   total signal counts    & $\sim$ 2600 counts   \\ \hline
      $R$                       &   Gaussian FWHM      & 0.65"  \\ \hline
      $n_{\rm gal}$      &   number density         & $\sim$ 5.5 $/ {\rm arcmin^{2}}$ \\ \hline
      \chihway{SNR}   &  \chihway{signal-to-noise ratio} & \chihway{8.33} \\ \hline 
  \end{tabular}
  \label{table:fiducial_gal}  
\end{table}

To further simplify the problem, the distribution of circular galaxies is collapsed into a single circular 
Gaussian shape. By exploring the size-magnitude parameter space, we find that using roughly the 
average magnitude, size and number density of the original sample, we can recover the ellipticity 
correlation of the circularised galaxy sample. For the rest of this paper, we will refer to this special 
circular Gaussian as the ``fiducial galaxy.'' \Fref{fig:nulltest_cf} shows the ellipticity correlation 
functions for three different sizes of circular Gaussian shapes, with the middle one (square) being the 
fiducial galaxy. The characteristics of the fiducial galaxy are listed in \Tref{table:fiducial_gal}. 

\chihway{The construction of the fiducial galaxy is an approximation, but is appropriate for our 
analyses with the following caveats. First, by taking the ellipticity results from a circular 
galaxy ($\boldsymbol{\varepsilon}^{i}=\boldsymbol{\gamma}=0$) as a general result for the whole 
galaxy population, we are assuming that the average ellipticity error on the population of galaxies is 
approximately the ellipticity error on the average galaxy in the population. This is ignoring the fact that 
certain algorithms may tend to measure the ellipticity of a galaxy more accurately when the galaxy is 
more circular or more elliptical. This intrinsic-ellipticity-dependent error may introduce additional errors 
in the ellipticity measurements. We ignore them because these errors are algorithm-dependent, and are 
spatially uncorrelated, \ie they only contribute a small addition contribution to shape noise.
Second, by choosing a Gaussian profile for the fiducial galaxy rather 
than a more realistic Sersic-type profile, we are assuming our ellipticity measurement method performs 
equally well on Gaussian profiles and realistic galaxy profiles. This again is to eliminate the algorithm-dependence 
coupling to the problem and also important later for shear measurements as discussed in 
\Sref{sec:Framework_ShearSys}.}      

\section{Quantifying errors on ellipticity measurements}
\label{sec:EllipSys}

In this section, if not otherwise specified, the measured ellipticity on any simulated galaxy image is 
effectively an ``ellipticity error'' generated from a certain physical effect, for reasons we have explained in 
\Sref{sec:Framework_circular_gaussians} ($\boldsymbol{\varepsilon}^{i}=\boldsymbol{\gamma}=0$). 
Thus we omit the superscripts in our notation and use $\boldsymbol{\varepsilon}$  
($\xi_{\varepsilon\varepsilon}$) instead of $\varepsilon^{s}$ ($\xi_{\varepsilon^{s}\varepsilon^{s}}$) 
or $\varepsilon^{m}$ ($\xi_{\varepsilon^{m}\varepsilon^{m}}$).
 
Also, for all ellipticity measurements, we define the quantity $\sigma[\boldsymbol{\varepsilon}]$ to be a 
measure of the uncertainty in ellipticity measurements due to a certain physical effect. 
$\sigma[\boldsymbol{\varepsilon}]$ is defined as the square-root of the quadrature sum of the standard 
deviation of individual $\varepsilon_{1}$ and $\varepsilon_{2}$ distributions (as opposed to the standard 
deviation of $\varepsilon$):

\begin{equation}
   \sigma[\boldsymbol{\varepsilon}]=\sqrt{\sigma_{\varepsilon_{1}}^{2}
  +\sigma_{\varepsilon_{2}}^{2}} \; .
\label{eq:sigmae}  
\end{equation}

\subsection{Non-stochastic effects}
\label{sec:EllipSys_NS}
 
\subsubsection{Simulations}
\label{sec:EllipSys_NS_sim}
To examine the non-stochastic effects, we generated two sets of simulations. Each set of simulations 
consists of one or more full LSST focal planes. In all of the simulations, since we need to ``turn off'' the 
atmospheric effects, we convolve the fiducial galaxies with a circular Gaussian before running the 
simulations to ensure that the observed galaxies have the same size as a fiducial galaxy observed 
under the fiducial observing conditions. \chihway{To suppress the contribution from counting statistics 
errors in the measurements, all objects are generated with high SNR at $\sim 160$. }


In the first set of simulations, we simulate the ``as designed instrument'' by including only the optics 
design, isotropic charge diffusion in the CCD detectors, and pixelisation. In the second set of simulations, 
we include non-stochastic perturbations to the optics due to solid body misalignments, surface perturbations 
of the major optics elements, and warping and misalignment of the individual sensors in the focal plane (see 
\Tref{table:optics_parameters} for approximate levels of the major perturbations). 
\chihway{Twenty focal-plane size images with different Gaussian random realisations of these 
errors are generated in order to capture the effects in a ``typical'' LSST observations.}
        
\subsubsection{Results}
\label{sec:EllipSys_NS_results}
 
\begin{figure*}  
  \begin{center}
     \subfigure[]{\includegraphics[height=3in]{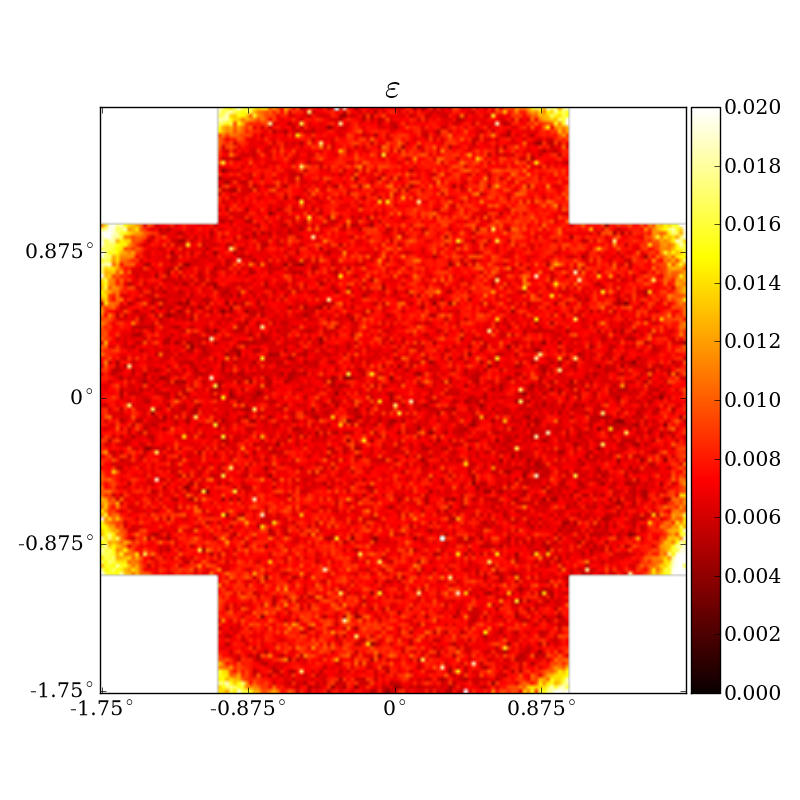}}
    \subfigure[]{ \includegraphics[height=3in]{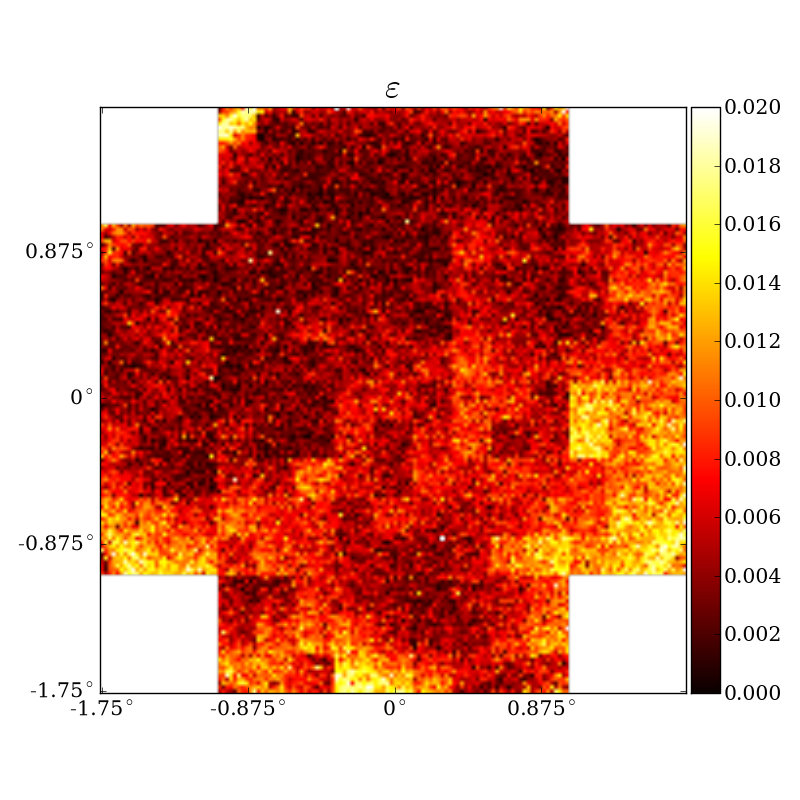}}
  \caption{Ellipticity magnitude $\varepsilon$ measured from the fiducial galaxies over the LSST focal 
  plane for the (a) optics design and (b) one example of adding non-stochastic optics errors. All 
  non-stochastic optics effects induce ellipticity magnitudes $<$0.02 for most of the field. Greater ellipticites 
  are mainly near the edges. The large-scale variations come from perturbations of the positions and 
  surface heights of the mirrors and lenses, while the variation in the focal plane height contributes to the 
  visible boundaries between individual CCDs.}
\label{fig:ns_maps}  
\end{center}
\end{figure*}

In \Fref{fig:ns_maps}(a), we show the ellipticity magnitude $\varepsilon$ measured from the 
simulations across the LSST focal plane for the as designed instrument. When non-stochastic 
optics error is induced, the measured ellipticity changes. One example in the set of simulation is shown in 
\Fref{fig:ns_maps}(b), where the change in ellipticity due to a certain set of optics errors is shown.        
 
\begin{figure*}  
  \begin{center}
   \subfigure[]{\includegraphics[height=2.4in]{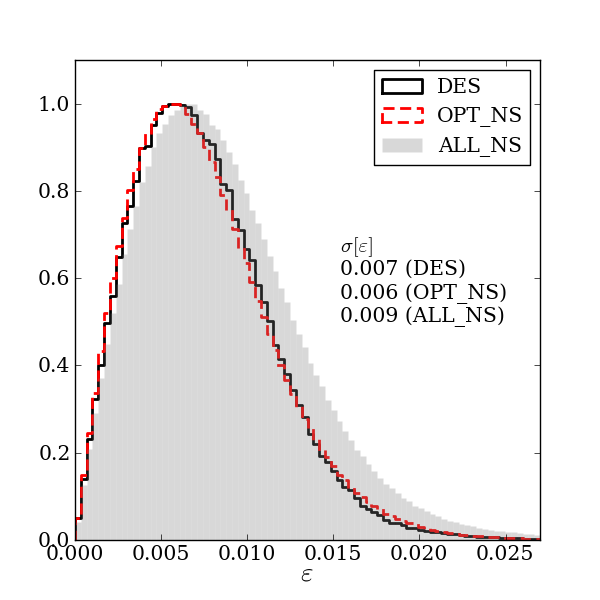}}
   \subfigure[]{\includegraphics[height=2.4in]{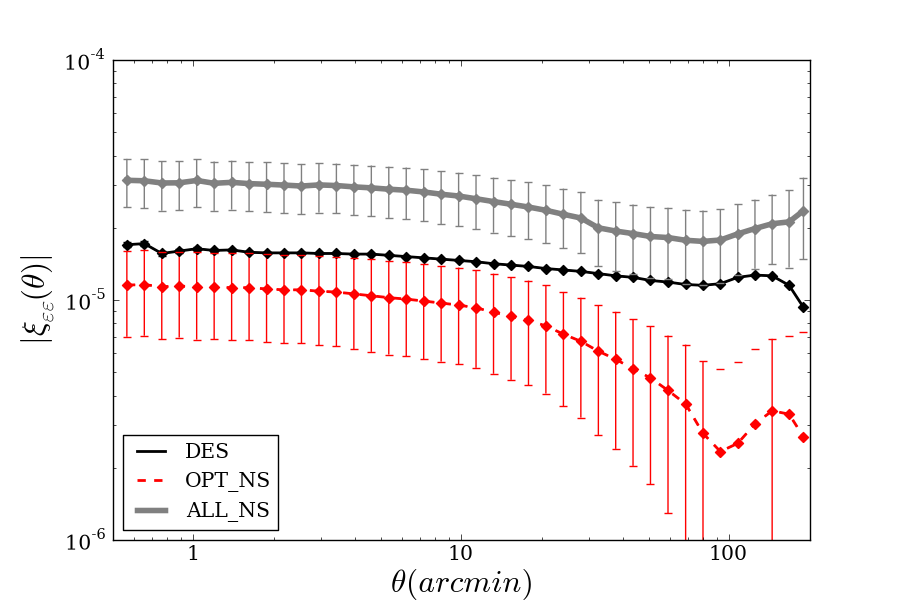}}
  \caption{The following abbreviations are used for the different sources of ellipticity errors: DES 
  (design), OPT\_NS (non-stochastic optics effects) and ALL\_NS (all non-stochastic effects). 
  (a) Distribution of the ellipticity magnitude measured for the fiducial galaxies when different 
  non-stochastic effects are added. (b) Absolute correlation function of the ellipticity errors for the fiducial 
  galaxies when different non-stochastic effects are added. The red curve is the median value for 20 
  different realisations of the non-stochastic optics effects.}
  \label{fig:ns_stats}  
  \end{center}
\end{figure*}

The distribution of $\varepsilon$ for the two sets are plotted in 
\Fref{fig:ns_stats}(a) with arbitrary normalisations. The corresponding $\sigma[\boldsymbol{\varepsilon}]$ 
values of these distributions after correcting for the counting statistics\footnote{\chihway{We will 
see later in \Sref{sec: EllipSys_scaling} that for the ellipticity measurements done with SNR$\sim$160 objects, 
there exist uncertainties from counting statistics at the $\sigma[\boldsymbol{\varepsilon}]\sim5\times10^{-3}$ 
level (see \Eref{eq:snr_fit}), which we need to subtract in quadrature from the raw measurements to isolate 
the uncertainties due to the specific effect of interest.}} are listed on the plot. We measure 
$\sigma[\boldsymbol{\varepsilon}]\sim7\times10^{-3}$ for the design, 
$\sigma[\boldsymbol{\varepsilon}]\sim6\times10^{-3}$ for the non-stochastic optics 
effects. The total ellipticity contribution from all non-stochastic effects on the fiducial galaxy is  
$\sigma[\boldsymbol{\varepsilon}]\sim 9\times10^{-3}$.
        
The median absolute ellipticity correlation function $|\xi_{\varepsilon\varepsilon}|$ measured from the 
fiducial galaxies in all the simulations is plotted in \Fref{fig:ns_stats}(b) for the design and for the 
non-stochastic optics effects added. The total ellipticity correlation function for all effects in this 
``non-stochastic'' class is also shown. The correlation function for the design is at the level 
$\sim 1.5\times10^{-5}$ with a rather flat shape. Adding non-stochastic effects almost doubles the 
level of the correlation function.    
        
Note that \Fref{fig:ns_stats} may not be characteristic of other telescopes, since we have utilised a large 
amount of LSST-specific information about the optics configuration and engineering tolerances 
\citep{SRD}. However, the main message from this section is the demonstration that for future large 
telescopes with designs similar to LSST, the non-stochastic spurious ellipticity correlation will be at a 
low level compared with existing telescopes \citep[see][for example]{2004astro.ph.12234J}. 

\subsection{Stochastic effects}
\label{sec:EllipSys_S}

\subsubsection{Simulations}      
\label{sec:EllipSys_S_sim}
        
To examine the contributions of stochastic effects, we use one realisation of the non-stochastic optics 
errors in the previous simulations and then add on stochastic contributions that vary randomly from 
exposure to exposure. Non-stochastic contributions to the ellipticities are later subtracted 
component-wise from the measured ellipticities to obtain the stochastic contribution. 

For each of the four stochastic effects listed in \Tref{table:classification}, we generate a set of 
20 focal-plane-size simulations with fiducial galaxies distributed over the field. The input 
parameters to the 20 simulations in each set are controlled so that each of the other three effects are 
``turned off'' and only one effect is ``turned on'' -- only parameters associated with that one effect 
are allowed to vary. In addition, we generate one set of simulations (20 focal-plane-size image), 
where the four stochastic effects are all turned on. These images are used in \Sref{sec:ShearSys} 
for shear measurement tests. We describe below the prescriptions for how we set the parameters for 
each of the effects in the simulations. 


\paragraph*{Counting statistics} is the largest stochastic source of noise in these single exposures. 
It is also the only effect that is stochastic in both space and time, which prohibits it from being 
corrected through PSF modelling.
        
The relevant measure for counting statistics is the SNR of an object, which we can calculate 
straightforwardly for a circular Gaussian profile given the total signal counts $S$, object FWHM 
size $R$, background counts $B_{\rm sky}$ and apparent object FWHM size $R^{m}$:
\begin{equation}
  {\rm SNR}=\frac{0.7\times S}{\sqrt{0.7\times S+\pi(1.34 \times R^{m})^{2}\times B_{\rm sky}}} 
\label{eq:snr}
\end{equation}
Here we use a typical aperture radius of $\sim$1.34 times the FWHM of the apparent 
object size $R^{m}$, containing $\sim$70\% of the source counts. For circular Gaussian, 
the apparent object size $R^{m}$ can be approximated as the object size $R$ convolved with a 
circular Gaussian with FWHM size equal to the PSF size $a$:
\begin{equation}
(R^{m})^{2}=R^{2}+a^{2}
\label{eq:Rm}
\end{equation}
\noindent where $a$ can be estimated by adding to $a_{\rm atm}$ in quadrature the 
instrumental PSF contribution $a_{\rm opt}$:
\begin{equation}
a=\sqrt{a_{\rm atm}^{2}+a_{\rm opt}^{2}}
\end{equation}

In this section, we fix $S$, $R$, $B^{\rm sky}$ and $a$ to the fiducial values in 
\trefa{table:fiducial_params} and \trefb{table:fiducial_gal}. In \Sref{sec: EllipSys_scaling}, 
we explore the full SNR parameter space. The 20 focal-plane-size images in this set are 
identical except that different photons are drawn from the galaxy and therefore travel 
slightly different light-paths. 
        
\paragraph*{Stochastic optics effects} are the residual optics errors after AOS correction that do 
not show repeatable patterns from exposure to exposure. The 20 focal-plane-size images in this 
set are identical except that the optics perturbations are varied from frame to frame within the level 
allowed by the adopted tolerances listed in \Tref{table:optics_parameters}. 
                
\paragraph*{Tracking errors} occur due to imperfect tracking of the telescope during the exposure. 
They cause the measured object shape to be slightly elongated in the direction of the sky rotation. 
The 20 focal-plane-size images are identical except that a different tracking error trajectory within 
the adopted tolerances described in \Sref{sec:ImSim_model_tracking} is assigned to each realisation. 
        
\paragraph*{Atmospheric effects} are slightly more complicated to model. For the same assumed 
fiducial seeing in all 20 simulations, different realisations of the atmosphere are generated by 
different combinations of the structure function, outer scale, wind speed and wind directions over the 
multiple atmospheric layers. The 20 focal-plane-size images are identical except that a different 
combination of these parameters is used. 
       
\vspace{0.2in}

For the first three sets of simulations, since the atmosphere is turned off, all objects are convolved 
with a circular Gaussian before we run the simulator so that the measured object size is the same as if 
it had propagated through the atmosphere. \chihway{Similar to \Sref{sec:EllipSys_NS_sim}, for all sets 
except the first, objects are generated with high SNR at $\sim 160$ to suppress contribution from 
counting statistics errors in the measurements.} The ellipticity of each object is measured and the 
mean ellipticity for each object over the 20 realisations is taken as the non-stochastic contributions and 
subtracted component-wise to yield the stochastic ellipticity component. 
 
\subsubsection{Results}
\label{sec:EllipSys_S_results}
 
\begin{figure*}  
\begin{center}
   \subfigure[]{\includegraphics[height=3in]{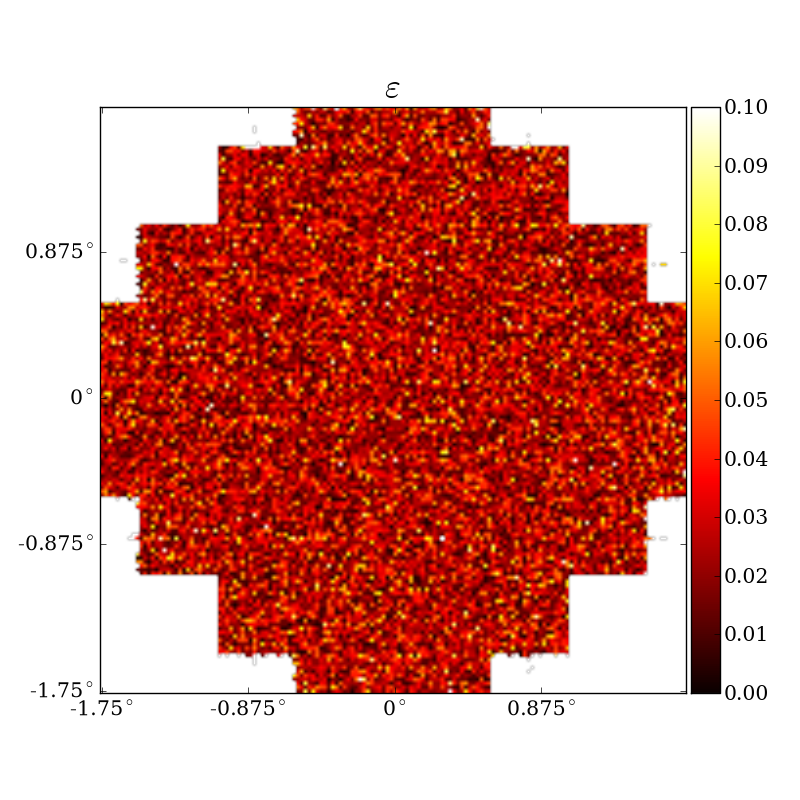}} 
   \subfigure[]{ \includegraphics[height=3in]{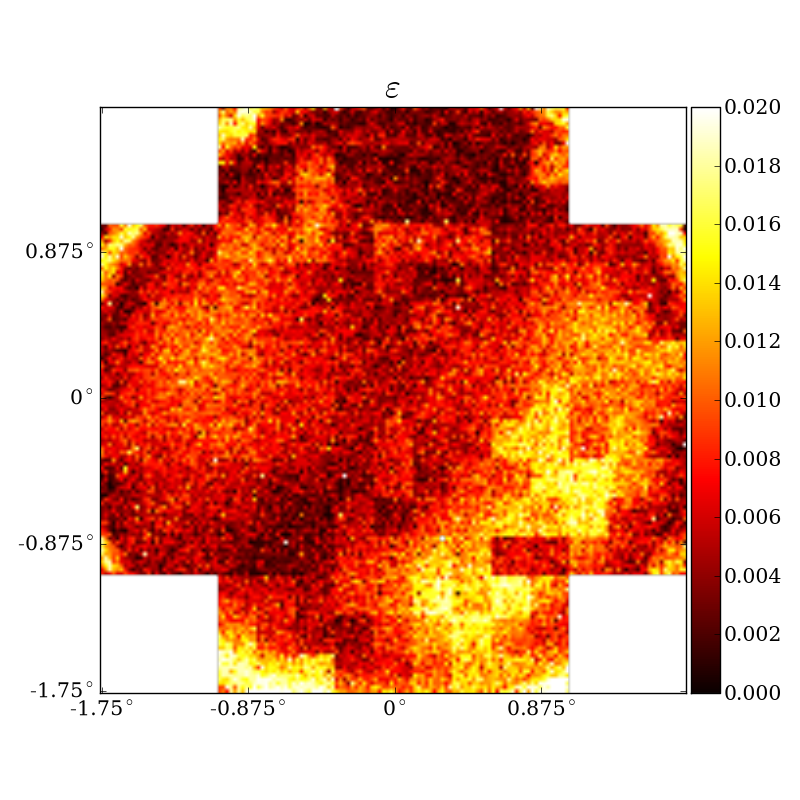} }\\
   \subfigure[]{\includegraphics[height=3in]{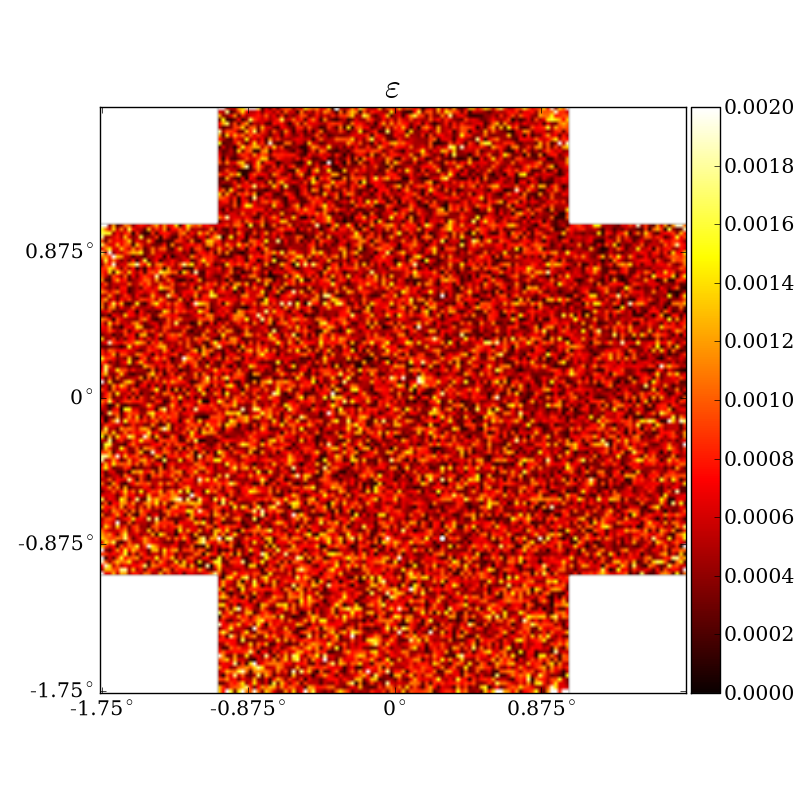}}
   \subfigure[]{\includegraphics[height=3in]{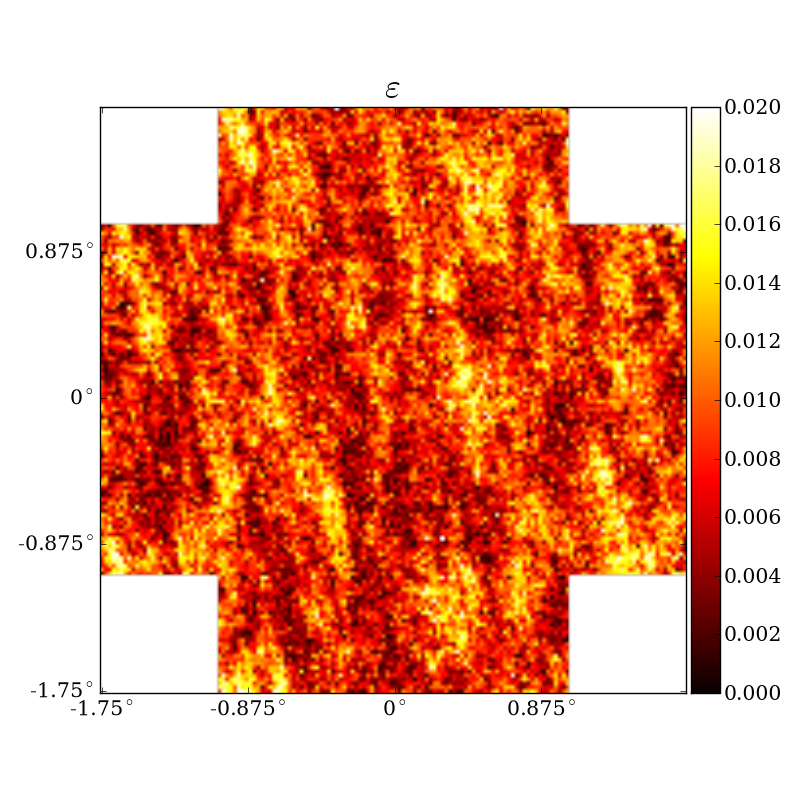} }
   \caption{Ellipticity magnitude $\varepsilon$ measured from the fiducial galaxies over the LSST 
   focal plane for four stochastic effects: (a) counting statistics, (b) tracking errors, (c) stochastic optics 
   errors and (d) atmospheric distortions. In each case, the colour bars are adjusted to best show the 
   ellipticity spatial pattern. Since these effects are stochastic, we show only one representative 
   realisation of the random process to illustrate the kind of ellipticity pattern induced by each effect. 
   Among the four, (a) counting statistics induces the highest level of errors and show ``missing'' 
   sensors on the edge of the field due to the fact that vignetting causes the fiducial galaxies to be 
   undetectable at those positions.}
\label{fig:s_maps}  
\end{center}
\end{figure*}

\Fref{fig:s_maps} shows one example of the absolute ellipticity errors in each of the four sets. 
The colour mapping for the four plots are adjusted to best illustrate the spatial patterns and absolute 
ellipticity levels. Note that in \Fref{fig:s_maps}(a) the CCDs in the corner of the field are missing. 
This is because the fiducial galaxies have very low SNR at those vignetted locations. In a more 
realistic field, brighter galaxies will still be detected there.  

\begin{figure*}  
\begin{center}
  \subfigure[]{\includegraphics[height=2.4in]{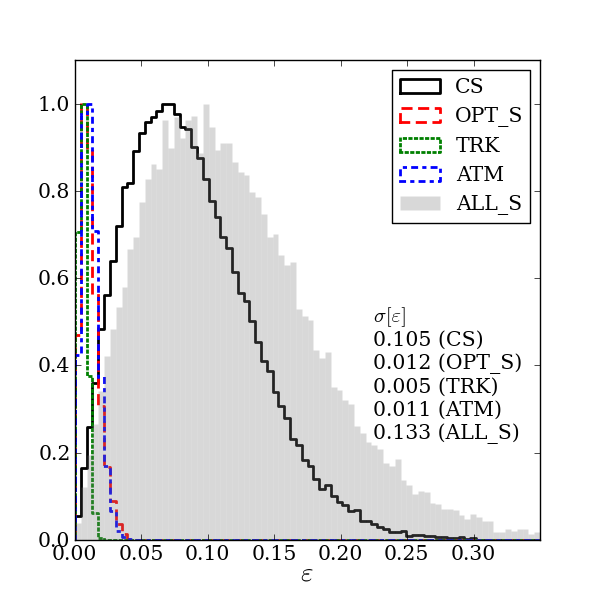}} 
  \subfigure[]{\includegraphics[height=2.4in]{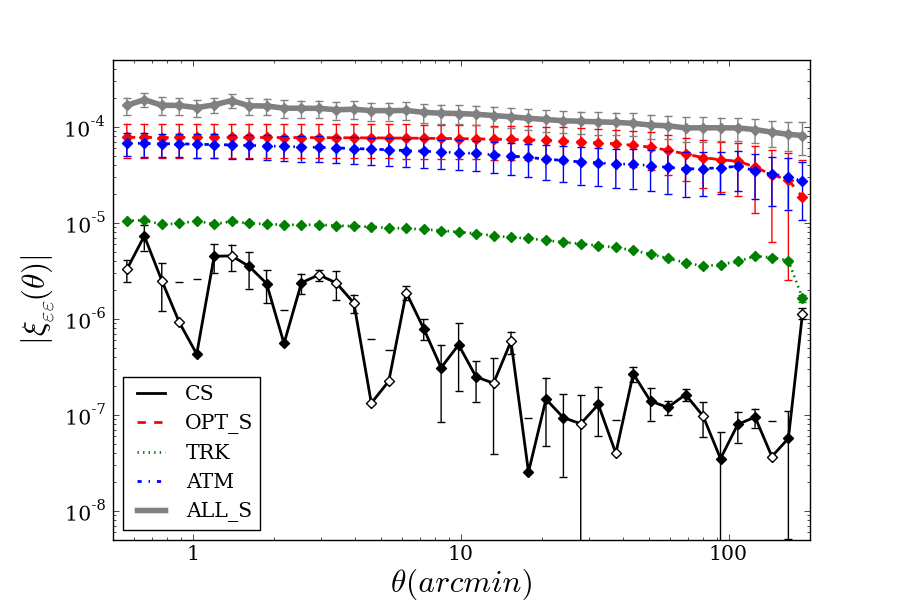}} \\
  \subfigure[]{\includegraphics[height=2.4in]{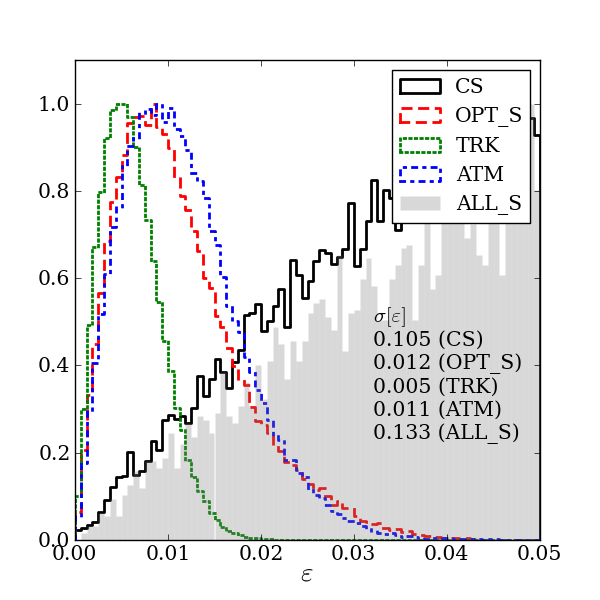}} 
  \subfigure[]{ \includegraphics[height=2.4in]{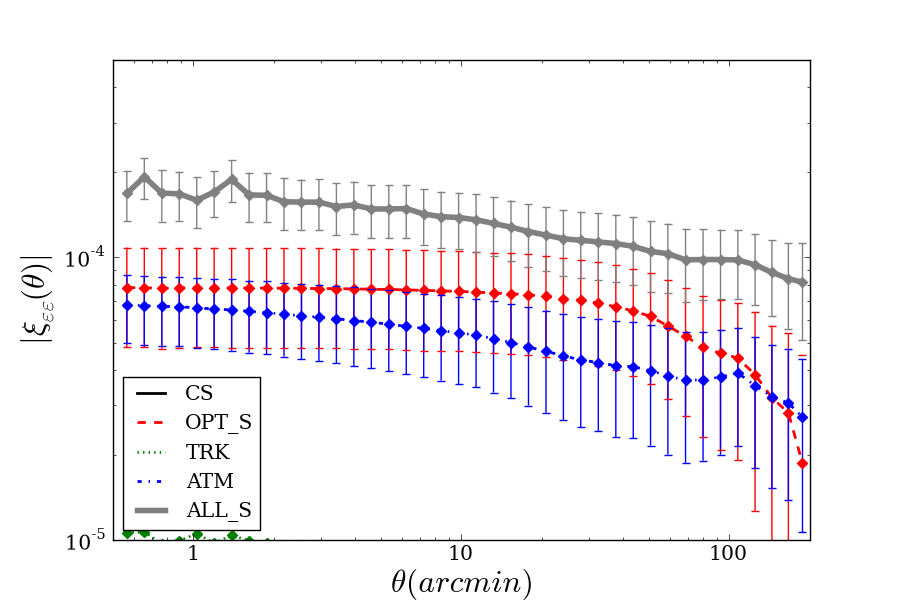}}
  \caption{The following abbreviations are used for the different sources of ellipticity errors: CS (counting 
  statistics), OPT\_S (stochastic optics effects), TRK (tracking errors), ATM (atmospheric effects), and 
  ALL\_S (all stochastic effects). (a) Distribution of the ellipticity magnitude measured for the fiducial 
  galaxies when different stochastic effects are added. The grey shaded area indicates the ellipticity magnitude 
  distribution from all the stochastic effects together. (b) Absolute correlation function of the ellipticity errors for 
  the fiducial galaxies when different stochastic effects are added. Note that all curves plotted for the stochastic 
  effects are the median value for 20 different realisations, with the error bars showing the standard deviation 
  of the 20 realisations divided by $\sqrt{20}$. Negative values are plotted with open symbols. (c) is a 
  zoomed-in view of (a) on the lower ellipticity values and (d) is a zoomed-in view of (b) on the higher 
  correlation function curves.}
\label{fig:s_stats}  
\end{center}
\end{figure*}

Distributions of the magnitudes of the stochastic ellipticity errors $\varepsilon$ measured from the four sets 
of simulations are plotted in \Fref{fig:s_stats}(a). Each of the curve is normalized so that it peaks at . Also overlaid 
is the total stochastic ellipticity error distribution. \chihway{The corresponding $\sigma[\boldsymbol{\varepsilon}]$ 
values of these distributions after correcting for counting statistics$^{11}$ are listed on the 
plot.} \Fref{fig:s_stats}(c) is a zoomed-in view of \Fref{fig:s_stats}(a) on the lower ellipticity values. Clearly, 
in a single exposure the dominant error contribution to the shape measurements for a fiducial galaxy 
is counting statistics, giving $\sigma[\boldsymbol{\varepsilon}]\sim 1.1\times10^{-1}$. The atmospheric 
effects and the stochastic optics effects are at similar levels and are the second and third largest 
contributors, giving $\sigma[\boldsymbol{\varepsilon}]\sim 1.2\times10^{-2}$ 
and $\sigma[\boldsymbol{\varepsilon}]\sim 1.1\times10^{-2}$ 
respectively. Tracking errors are the most insignificant effect of the four, with 
$\sigma[\boldsymbol{\varepsilon}]\sim 5\times10^{-3}$. 
The total stochastic ellipticity uncertainty when all effects are turned on is 
$\sigma[\boldsymbol{\varepsilon}]\sim 1.3\times10^{-1}$. Note that the total non-stochastic ellipticity error 
discussed in \Sref{sec:EllipSys_NS} is more than an order of magnitude smaller than the total stochastic 
ellipticity errors. 

We now turn to the median correlation functions of these four stochastic ellipticity error components, 
shown in \Fref{fig:s_stats}(b), along with the total stochastic ellipticity error correlation function. The error 
bars in each case show the standard deviation of the 20 realisations divided by $\sqrt{20}$. The first 
observation is that although counting statistics errors dominate in the ellipticity error as shown 
in \frefa{fig:s_maps} and \frefb{fig:s_stats}(a), they are completely uncorrelated. They oscillate rapidly but are 
always consistent with zero. This is of course expected -- regardless of the SNR of 
the measured objects, counting statistics makes no contribution to the correlation function. Tracking errors, on 
the contrary, being the lowest in \Fref{fig:s_stats}(a), contribute to a small but non-zero correlation at 
the $10^{-5}$ level. The stochastic optics effects generate ellipticity correlations slightly below the $10^{-4}$ 
level while the atmospheric effects contribute to a  similar level of ellipticity correlation with a 
steeper shape as can be seen more clearly in \Fref{fig:s_stats}(d), the zoomed-in view for \Fref{fig:s_stats}(b). 
The total ellipticity correlation function is thus dominated by the atmosphere component at small scales and 
then a combination of the stochastic optics errors and the atmospheric effects at larger scales. Also notice that 
the non-stochastic component is approximately an order of magnitude smaller than the total stochastic ellipticity 
correlation in a single 
exposure. 

\subsection{Discussion}
\label{sec: EllipSys_scaling}

Up to this point, we have quantified $\sigma[\boldsymbol{\varepsilon}]$, the expected levels of errors 
in ellipticity measurements due to different physical effects for a typical LSST single exposure. These 
results can be scaled to other observing conditions and source distributions as a first-order estimation 
for the uncertainties in ellipticity measurements in another dataset. The two major quantities that govern 
the scaling of $\sigma[\boldsymbol{\varepsilon}]$ and $\xi_{\varepsilon\varepsilon}$ for a certain 
dataset are the average observed object size $R^{m}$ (\Eref{eq:Rm}) and the average SNR 
(\Eref{eq:snr}) of the objects. The level of counting statistics contribution to 
ellipticity errors $\sigma[\boldsymbol{\varepsilon}]$ is expected to scale with some function of SNR, while 
ellipticity errors from all the other effects scale with\footnote{This can be derived by 
assuming the measured ellipticity comes from an elliptical Gaussian PSF convolved with a circular Gaussian 
galaxy. If the PSF has second moments $I'_{ij}$ and the galaxy has second moments $I_{ij}$, then because 
the moments are effectively summed in the convolved image, and $I_{11}-I_{22}=I_{12}=0$, the convolved 
object has ellipticity 
$(\varepsilon_{1},\varepsilon_{2})=\frac{1}{I_{11}+I'_{11}+I_{22}+I'_{22}}(I'_{11}-I'_{22}, 2I'_{12})
\propto(\frac{1}{R^{m}})^{2}(I'_{11}-I'_{22}, 2I'_{12})$.} $(R^{m})^{-2}$. For $\xi_{\varepsilon\varepsilon}$, on 
the other hand, the counting statistics contribution is essentially zero, while all other components scale 
with $(R^{m})^{-4}$. 

\chihway{Note also that in general the measured level of the correlation function is lower than what is 
expected for a naive assumption of $|\xi_{\varepsilon\varepsilon}| \sim \sigma[\boldsymbol{\varepsilon}]^2$. 
This is because the distortions of the galaxies are usually only partially correlated in space. The degree of 
correlation, which is governed by the physical mechanism that induces the correlation, determines how 
close $|\xi_{\varepsilon\varepsilon}|$ approaches $\sigma[\boldsymbol{\varepsilon}]^2$.} 

\begin{figure}  
\begin{center}
  \includegraphics[height=2.6in]{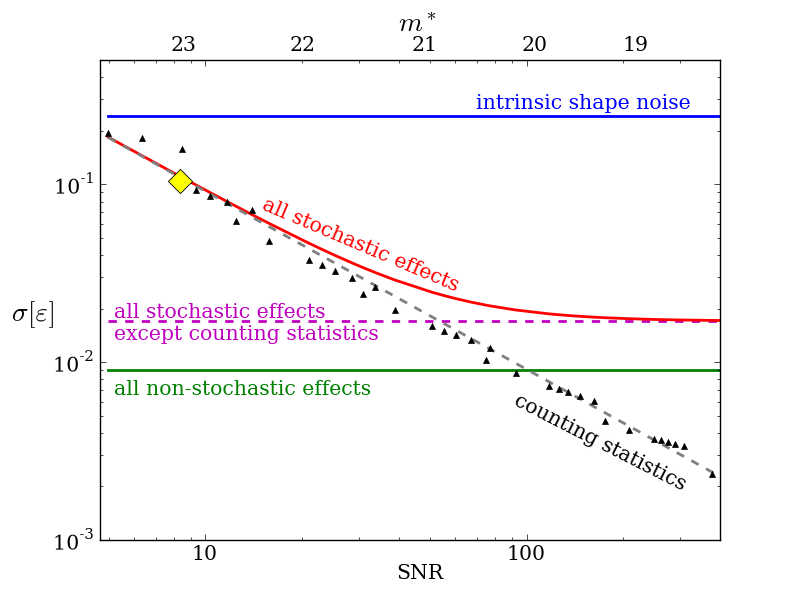}
  \caption{Uncertainties of ellipticity measurements induced by counting statistics errors for objects in a 
  large range of size and magnitudes is labeled by the black triangles, which is well fit by 
  a function of the objects' SNR as shown with the grey dashed line. If all the parameters other than 
  magnitude ($m$, or effectively, $S$) are held fixed, then the bottom SNR axis corresponds to the $r$-
  band AB magnitudes labeled on the top axis $m^{*}$. The yellow diamond at $m^{*}=23$ is where the 
  fiducial galaxy is located. Under the same assumption that we only vary the object's magnitude along 
  the x-axis, we can then plot the ellipticity uncertainties induced by all the other physical effects discussed 
  in this paper as horizontal lines since these errors depends only on the object's size. Finally, the level of 
  intrinsic shape noise is indicated by the solid blue line.}
\label{fig:snr}  
\end{center}
\end{figure}

To determine the scaling of $\sigma[\boldsymbol{\varepsilon}]$ with SNR, we perform a series of 
simulations similar to the first set of simulations (\ie counting statistics) in 
\Sref{sec:EllipSys_S_sim}, but vary the input galaxy's size and magnitudes over the range $R=$ [0" (point 
source), 0.5", 0.7", 0.85", 1.0", 1.5", 2.0"] and $m=$ [18, 19, 20, 21, 22, 23, 24] to cover a nominal galaxy 
population and plot $\sigma[\boldsymbol{\varepsilon}]$ as a function of the object's SNR. The results are 
illustrated in \Fref{fig:snr}. We find that even for the wide range of size and brightness sampled, the ellipticity 
errors for all objects lie on a power law curve of index $\sim-1$, described by the fit: 
\begin{equation}
  \sigma[\boldsymbol{\varepsilon}_{CS}]\approx0.875\times {\rm SNR}^{-0.9995}\approx \frac{0.875}{\rm SNR}\;,
  \label{eq:snr_fit}
\end{equation}

\noindent where the subscript $CS$ indicates the ellipticity uncertainty due to counting statistics errors only.
\chihway{\Eref{eq:snr_fit} is consistent with the analytical predictions and numerical simulations 
\citep{2008A&A...484...67P, 2012arXiv1203.5050R} in previous studies. }
   
In \Fref{fig:snr}, we show the breakdown of $\sigma[\boldsymbol{\varepsilon}]$ into different 
components for a galaxy of FWHM size $\sim0.65$" under fiducial LSST observing conditions 
(\Tref{table:fiducial_params}) as a function of the SNRs of the objects. The top axis $m^{*}$ shows the 
corresponding $r$-band AB magnitude for objects at that SNR. The errors due to all non-stochastic effects 
and the errors due to all stochastic effects except counting statistics are by definition independent of the 
SNR of the galaxy, therefore they are represented by horizontal lines on the plot. The total errors from 
stochastic effects are derived by adding the level of counting statistics contributions and other stochastic 
errors in quadrature. 


Under the assumption that these individual noise terms are approximately independent from one another, 
we can now estimate the uncertainty in ellipticity measurements of an arbitrary galaxy under an arbitrary 
scenario by scaling the results from our tests with the fiducial galaxies and conditions, which we denote by the 
subscript ``0'' in the following steps:
 
\begin{enumerate}
  \item Calculate the galaxy's SNR, which depends on $R^{m}$, $S$, $a$ and $B_{\rm sky}$, and scale the 
  counting statistics errors on the ellipticity from the fiducial case via:
  \begin{equation}
    \sigma[\boldsymbol{\varepsilon}]\propto \frac{SNR_{0}}{SNR}\;.
  \label{eq:scaling1}
  \end{equation}  
  \item Calculate the contributions from all other effects by scaling the results in \srefa{sec:EllipSys_NS} and 
  \srefb{sec:EllipSys_S} by the measured galaxy size:
  \begin{equation}
    \sigma[\boldsymbol{\varepsilon}]\propto \left(\frac{R^{m}_{0}}{R^{m}}\right)^{2}\;.
  \label{eq:scaling2}
  \end{equation}  
   \item All stochastic components are further scaled by the exposure time:
   \begin{equation}
     \sigma[\boldsymbol{\varepsilon}]\propto\sqrt{\frac{t_{exp,0}}{t_{exp}}}\;.
   \label{eq:scaling3}
   \end{equation}  
   \item Add the individual components in quadrature to yield an estimate of the total ellipticity uncertainty 
   in the measurement. Note, however, that the simple assumption that all effects are decoupled breaks 
   down at the low SNR end, where the errors are no longer small and cannot be linearly decomposed 
   into the different components. 
\end{enumerate}
      
It is straightforward to also estimate the ellipticity error correlation function of a population of galaxies:

\begin{enumerate}
   \item Since counting statistics errors do not correlate, we do not need to account for a correlation function 
   for them.  
   \item Scale the individual correlation functions in \frefa{fig:ns_stats} and \frefb{fig:s_stats} by the average 
   $R^{m}$ of objects in the frame:
  \begin{equation}
    \xi_{\varepsilon\varepsilon}\propto \left( \frac{R^{m}_{0}}{R^{m}} \right)^{4}\;.
  \label{eq:scaling4}
  \end{equation}     
   \item Scale the stochastic components of the individual correlation functions by $t_{exp}$: 
  \begin{equation}
    \xi_{\varepsilon\varepsilon}\propto \frac{t_{exp,0}}{t_{exp}}\;.
  \label{eq:scaling5}
  \end{equation} 
   \item Add the individual components to yield the total ellipticity correlation function.
\end{enumerate}

These scaling relations serve as first-order estimates of the ellipticity errors and error 
correlations.

\section{Sources of spurious shear}
\label{sec:Framework_ShearSys}
 
After measuring the ellipticities of the galaxies ($\boldsymbol{\varepsilon}^{m}$ in \Eref{eq:measured_e}), 
the next step, following \Eref{eq:measured_g}, is to estimate the PSF-induced ellipticity errors 
$\boldsymbol{\varepsilon}^{s}$, estimate the scaling factor $\alpha$ and calculate the shear 
estimator $\boldsymbol{\hat{\gamma}}$:
\begin{equation}
  \boldsymbol{\hat{\gamma}}
  =\frac{\alpha}{2}(\boldsymbol{\varepsilon}^{m}-\boldsymbol{\varepsilon}^{s})
  =\boldsymbol{\gamma}+\frac{\alpha}{2}\boldsymbol{\varepsilon}^{i}.
\label{eq:measured_g}  
\end{equation}
The first part of \Eref{eq:measured_g} represents operationally how one would calculate 
$\boldsymbol{\hat{\gamma}}$ and the second part comes from rearranging \Eref{eq:measured_e}.
When averaging $\boldsymbol{\hat{\gamma}}$ over a large number of galaxies, we recover the 
true shear, or $\langle \boldsymbol{\hat{\gamma}} \rangle= \boldsymbol{\gamma}$. As described in 
\Sref{sec:Framework_circular_gaussians}, since we have set both shear and intrinsic ellipticity 
to zero in our simulations, any non-zero shear measurement from \Eref{eq:measured_g}, even 
without averaging over an ensemble of galaxies, indicates  
mis-estimation of $\boldsymbol{\varepsilon}^{s}$ and/or $\alpha$. In other words, any measured 
shear from our simulations is \textit{spurious}. In the remainder of this section, we use 
$\boldsymbol{\gamma}^{s}$ and $\xi_{\gamma^{s}\gamma^{s}}$ to indicate the spurious shear and 
their correlation functions. As opposed to previous ellipticity measurements, we only show results for 
$\xi_{\gamma^{s}\gamma^{s}}$ (rather than single $\boldsymbol{\gamma}^{s}$ measurements), and 
propagate them directly into uncertainties of the inferred cosmological model in \Sref{sec:Cosmology}.

Operationally, three separate steps are involved in calculating the terms of \Eref{eq:measured_g} that 
are prone to systematic effects: PSF modelling (estimating $\boldsymbol{\varepsilon}^{s}$), 
``deconvolving'' with the PSF\footnote{Deconvolution here implies some algorithm that removes the 
effects of the PSF from the galaxy images, which may or may not be a mathematically exact 
deconvolution.} 
(properly removing the effect of the PSF in estimating intrinsic galaxy shape) and converting the 
measurement to shear (estimating $\alpha$). The first step is related to properly modelling the physical 
effects discussed previously, while the latter two steps are determined by choices of algorithms. We 
use the terms ``spurious shear from PSF modelling'' and ``spurious shear from shear 
measurement algorithms'' to refer to these two classes of errors. 

\chihway{
Given a perfectly known PSF, an imperfect algorithm can still render spurious shear. Several works 
have studied the effectiveness of different PSF deconvolution algorithms and quantified the errors in 
shear measurements \citep{2006MNRAS.368.1323H,2007MNRAS.376...13M, 2010MNRAS.405.2044B, 
2012MNRAS.423.3163K, 2012arXiv1204.4096K}. This spurious shear studied in previous work 
(``spurious shear from shear measurement algorithms'' in our classification) is not necessarily intrinsic 
to the measurement, and is therefore not the main interest of this paper. We choose to use KSB, one of 
the most popular weak lensing algorithms, as our test method, but design the simulations and analyses 
as described below to eliminate some of the known flaws in this method. All of our results can 
thus be viewed as the \textit{best possible} results achievable by a KSB pipeline. In principle, more 
sophisticated pipelines should do even better.}

\chihway{
As mentioned in \Sref{sec:Framework_circular_gaussians}, we choose to use simple circular 
Gaussians as galaxies to perform our analyses. In terms of shear measurement, this implies that 
our estimates of the spurious shear from these galaxies will not be heavily affected by the choice 
of a simplistic moment-based method like KSB. Furthermore, to account for the ``calibration 
factors'' often used in KSB-like algorithms \citep{2006MNRAS.368.1323H,2007MNRAS.376...13M, 
2010MNRAS.405.2044B, 2012MNRAS.423.3163K, 2012arXiv1204.4096K}, 
which are derived from simulations and intended to calibrate the process that converts 
ellipticity to shear, we use a calibration factor that shifts the mean shear in each frame to zero, 
which effectively performs a perfect calibration for the additive shear error.}    

On the other hand, spurious shear induced by PSF modelling is less dependent on the specific 
shear measurement algorithm; instead, it is heavily affected by the nature of the various physical 
effects. In fact, to model the PSF across an image, we are really just modelling the response function 
of a point source to all the physical effects across the focal plane. For a multi-epoch survey like LSST, 
the two classes of physical effects -- non-stochastic and stochastic -- should be modeled differently. 

For the non-stochastic errors, since they show repeated patterns over multiple exposures, there is 
a massive number of stars that contain information to constrain the model. \citet{2004astro.ph.12234J} 
first suggested the concept of detecting the repeated patterns in the data themselves via principle 
component analysis (PCA). For current surveys, this is becoming a standard operation for PSF modelling 
in weak lensing analyses. The power of PCA scales with the total number of stars in all the exposures, 
which essentially scales with $1/\sqrt{N}$, where $N$ is the number of exposures taken with similar 
observing configurations \citep{2006JCAP...02..001J}. For LSST, we believe that the large number of 
exposures in the survey would enable us to characterise the non-stochastic PSF very accurately. PSF 
variation induced by stochastic effects, however, can be captured only from stars in a single exposure, 
which are both sparse and noisy. The stochastic PSF variation would be modeled poorly in a PCA-like 
approach.

\chihway{Note that the two classes of shear measurement errors are not necessarily decoupled, making 
them difficult to separate from one another. In this work, our goal is to quantify the former, the ``spurious 
shear from PSF modelling''. We do this by first eliminating the algorithm-dependence in our shear 
measurement algorithm by carefully designing the simulations and analysis pipeline, and then by testing 
for any residual shear errors from the algorithms with perfect knowledge of the PSF model 
(\Sref{sec:ShearSys_all_psf}).}

\section{Quantifying errors on shear measurements}
\label{sec:ShearSys}

We have shown in \Sref{sec:EllipSys} that the total ellipticity error correlation is at the $10^{-4} -10^{-3}$ 
level for a fiducial LSST single exposure. In this section, we correct the PSF effects in these simulations 
and measure the spurious shear correlation. Three different PSF model scenarios are considered: 
The first assumes perfect knowledge of the PSF; the second assumes that a PCA-like method is 
used to model the PSF, yielding perfect knowledge of the non-stochastic component of the PSF but no 
information about the stochastic component of the PSF; the third assumes that we attempt to model both 
components of the PSF simultaneously by using a standard method -- interpolating a smooth polynomial 
function between measurements of individual stars. By performing these three tests and examining the 
residual shear correlation function, we can pin down the sources of spurious shear correlation functions.  

All analyses are measurements of the spurious shear from fiducial galaxies 
in the set of 20 focal-plane-size simulations described in \Sref{sec:EllipSys_S_sim} that contain 
all the physical effects modelled in \phosim; we will refer to this set of simulations as the 
``master set''. In the three subsections below, we describe the simulations used to obtain the 
three different PSF models and show the spurious shear correlation functions we measure 
from the master set using the three PSF models. 

Also, if not otherwise specified, since the measured shear of any simulated image is effectively 
``spurious shear'' generated from the PSF modelling and correction process 
($\boldsymbol{\varepsilon}^{i}=\boldsymbol{\gamma}=0$), we omit the superscripts in our notation 
and use $\boldsymbol{\gamma}$  ($\xi_{\gamma\gamma}$) instead of $\gamma^{s}$ 
($\xi_{\gamma^{s}\gamma^{s}}$) or $\hat{\varepsilon}$ ($\xi_{\hat{\varepsilon}\hat{\varepsilon}}$).

\subsection{Perfect PSF model}
\label{sec:ShearSys_all_psf}
In this test, since the spurious shear from PSF modelling is by definition zero, the spurious shear 
we measure indicates any imperfections of the KSB implementation we adopted. 

\subsubsection{Simulations and results}
\label{sec:ShearSys_all_psf_results}
We generate a set of 20 focal-plane-size images identical to the master set, except that at the 
location of each galaxy, we simulate a bright star instead. The shape of each bright star is 
measured and the shape parameters are used to construct the PSF models for its galaxy 
partner in the master set. 

In \Fref{fig:g_cf}(a), we show the median shear correlation function for the 20 simulations. We show that 
by using a perfect PSF model, the spurious shear correlation is noisy but consistent with zero. 
This suggests that our idealised KSB implementation corrects the PSF effects nearly perfectly. Also plotted 
in \Fref{fig:g_cf}(a) for comparison are the ellipticity correlation function for the galaxies and the ellipticity 
correlation function of the PSF model. The PSF spatial correlation prints through and is apparent in the 
shear correlation as can be seen by the similarities between the blue and red curves. The error bars show 
the standard deviation in the 20 realisations divided by $\sqrt{20}$. 

        
\subsection{Perfect non-stochastic PSF model}
\label{sec:ShearSys_ns_psf}
Next we assume that the non-stochastic component of the PSF can be characterised perfectly by analysing 
a large number of exposures via, for example, a PCA method. However, no attempt to model the stochastic 
component of the PSF has been made. In this case, one would construct PSF models that capture only the 
non-stochastic effects discussed in \Sref{sec:EllipSys_NS}. The measured shear correlation function in this 
test is the spurious shear from not modelling and correcting for the the stochastic PSF variations.

\subsubsection{Simulations and results}
\label{sec:ShearSys_ns_psf_results}
Similar to \Sref{sec:ShearSys_all_psf_results}, we generate one focal-plane-size images with only the 
non-stochastic effects in the master set included and replace each galaxy in the master set with a bright star. 
The shape of each bright star is measured and the shape parameters are used to construct the PSF models 
for its galaxy partner in the master set. 

In \Fref{fig:g_cf}(b), we show the median shear correlation function for the 20 simulations. With corrections 
only for the non-stochastic component of the shape of the PSF, the spurious shear correlation for a single 
15-second exposure is at the few times $10^{-4}$ level. Also plotted for comparison are the ellipticity correlation 
function for the galaxies and the ellipticity correlation function of the PSF model. The error bars show the 
standard deviation in the 20 realisations divided by $\sqrt{20}$. Since we have shown in \frefa{fig:ns_stats} 
and \frefb{fig:s_stats} that the level of the stochastic ellipticity error correlation function is more than one order 
of magnitude larger than that for the non-stochastic ellipticity errors, it is reasonable that there are large spurious 
shear correlations when we correct only for the non-stochastic effects. The main effect of the PSF correction in 
this scenario is to correct for the PSF size and the weighting factor -- very little PSF ellipticity spatial variation 
is corrected. 

\subsection{Model both non-stochastic and stochastic PSF via polynomial models}
\label{sec:ShearSys_star_psf}    
In \Sref{sec:ShearSys_ns_psf_results} we have shown that even when the non-stochastic PSF is 
corrected, there can still be large shear residuals in single exposures due to stochastic PSF effects. This 
motivates us to model both stochastic and non-stochastic PSF components simultaneously. One common 
approach is to fit certain shape parameters of stars across the individual CCD sensors with a low order 
polynomial function, with the underlying assumption that the PSF spatial variation is smooth on individual 
sensor scales. The shear correlation function determined from this test is a measure of the spurious shear 
arising from incorrectly modelling and correcting for the stochastic and non-stochastic PSF variations 
using polynomial PSF models constructed from stars.

\subsubsection{Simulations and results}        

We generate a set of 20 focal plane-size images identical to the master set, except that the fiducial galaxies 
are replaced by a realistic star sample obtained from the \phosim sky catalogue, randomly located over the 
field. On average each sensor-size image contains $120 -150$ stars used for PSF modelling 
(SNR$>$13). The shape of each star is measured and the shape parameters are interpolated with 
nth-order polynomials onto the locations of the galaxies to obtain the PSF model at the location of the 
galaxies in the master set. We tested for several n values and show only the best case (n=5) here. 
                
\begin{figure}  
  \begin{center}
     \subfigure[]{\includegraphics[height=2.4in]{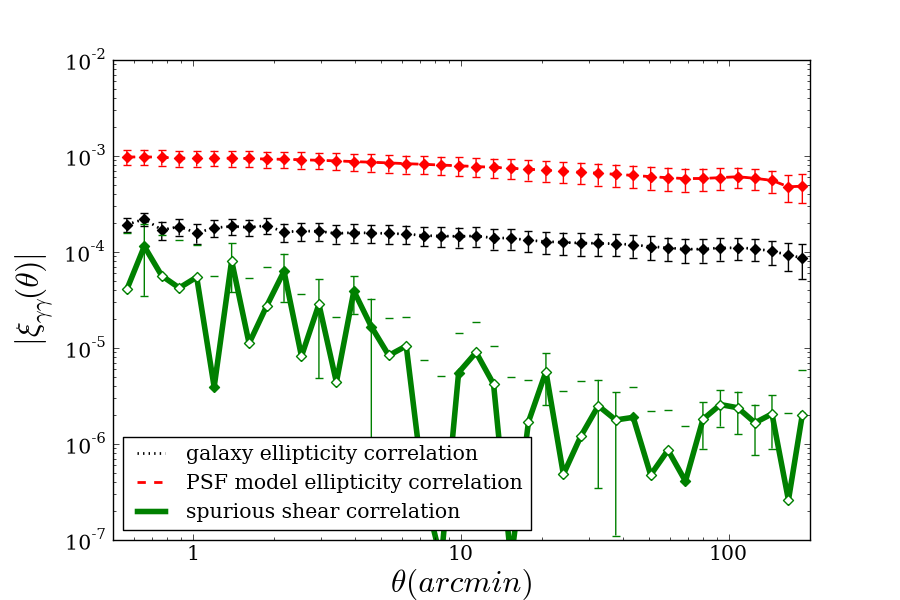}}	
     \subfigure[]{\includegraphics[height=2.4in]{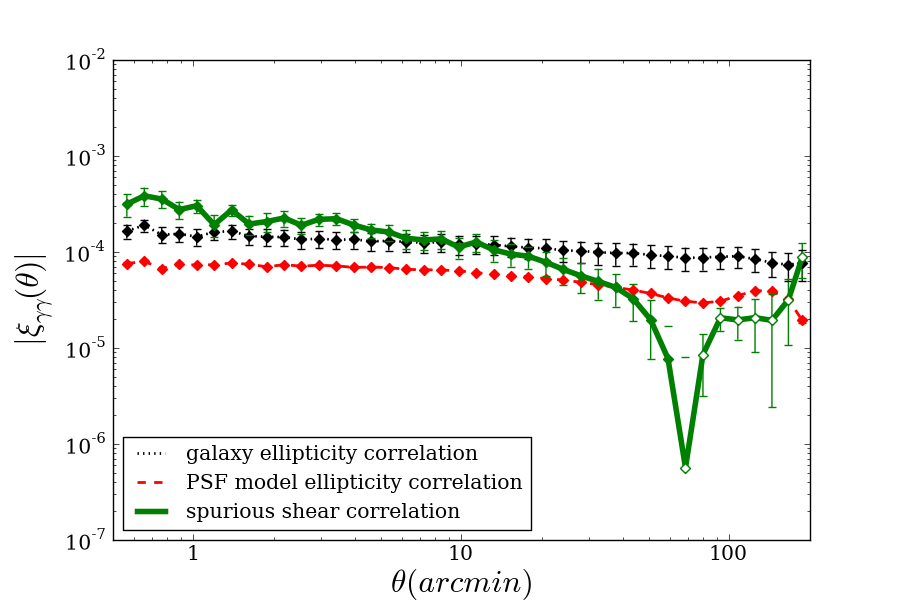}}
     \subfigure[]{\includegraphics[height=2.4in]{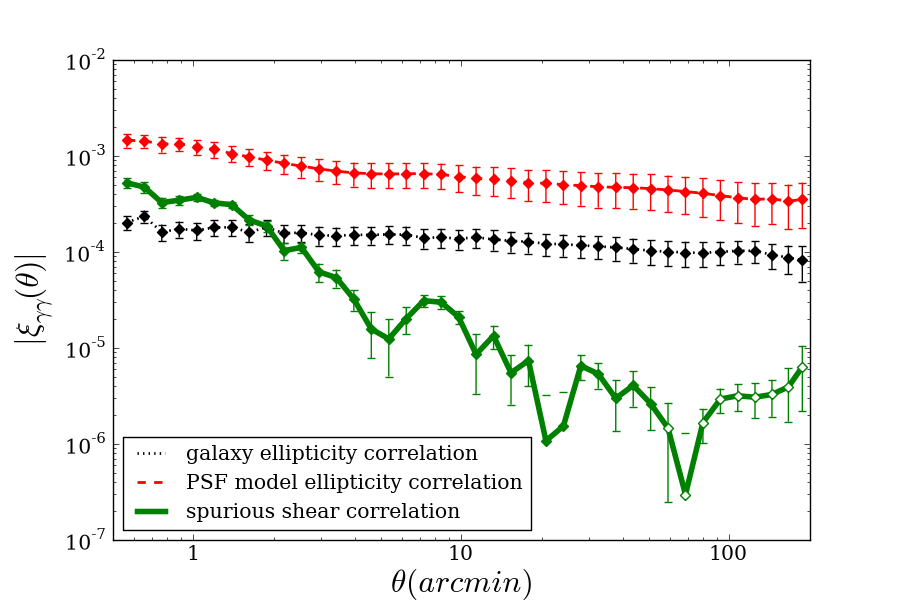} }
  \caption{The absolute ellipticity correlation function for the fiducial galaxies before PSF correction (dotted 
  black) and the absolute shear correlation function after PSF correction (solid green) for (a) perfect PSF 
  models, (b) perfect non-stochastic PSF models, and (c) PSF models constructed with a 5th-order polynomial 
  fit to bright stars. The absolute ellipticity correlation function (dashed red) in each case is plotted for 
  comparison. All curves are the medians of 20 different realisations under the fiducial observing condition, 
  with the error bars indicating the standard deviation in the 20 curves divided by $\sqrt{20}$. Negative values 
  are plotted with open symbols.}
  \label{fig:g_cf}  
  \end{center}
\end{figure}

\Fref{fig:g_cf}(c) shows the residual shear correlation functions when a 5th-order polynomial interpolation 
of stars is used to model the PSF. Also plotted are the ellipticity correlation function for the galaxies and the 
ellipticity correlation function of the 5th-order polynomial PSF model. The error bars show the standard deviation 
of the 20 realisations divided by $\sqrt{20}$.
        
Excess power is present on small scales in the shear correlation function and the slope has a slight transition  
at $\sim3'$, beyond which the curve decreases less steeply. 
\chihway{The negative correlation on large scales is an artifact from the shear calibration procedure described 
in \Sref{sec:Framework_ShearSys}, where the measured shear distribution in single measurements is calibrated 
to have zero mean, forcing part of the positive correlation to become negative. }
This excess power on small scales is expected, since structures within scales smaller than 
[sensor size]/n cannot be modeled by a polynomial of order n, where [sensor size]$\sim18'$ in our simulations, the 
part of the PSF not modeled by the polynomial prints through as spurious shear correlation. The fact that the PSF 
variations on small scales have significant power coming from the atmosphere, which we have shown in 
\Sref{sec:EllipSys_S}, means that the spurious shear correlation will also have excess power on these small scales. 
We measure the level of spurious shear correlation for a single 15-second exposure using a polynomial PSF model 
as $5\times10^{-4}$ at small scales and decreasing by two orders of magnitude towards larger scales. We have 
also examined how the different n values affect the level of the correlation function, and found that the general shape 
of the shear correlation remains similar to the n$=$5 case but the transition point where the correlation starts to rise 
at small scales changes according to [sensor size]/n. To improve upon this simple polynomial model, one would need 
to develop a more flexible interpolation technique that captures structures on different scales in a more efficient way.  
We propose such an approach in a companion paper \citep{C12}. 

\section{Discussion}
\label{sec:Discussion}

\subsection{Combining multiple exposures}
\label{sec:Combine}

\chihway{We now estimate the spurious shear correlation function in a combined 10-year LSST dataset. 
In the most simplistic case where all $N$ exposures on the same galaxy field have similar image 
quality, we show in \Aref{sec:Nscaling} that averaging the shear measurements in the $N$ exposures 
suppresses the stochastic piece of the spurious shear correlation by a factor $N$. But in a realistic case 
of varying image quality, the $N$ scaling is no longer straightforward. One needs to estimate the 
``effective number of exposures'', or $\Neff$, taken on each galaxy, which essentially weights each 
exposure according to the image quality. We direct the reader to \Aref{sec:Neff} for how we estimated 
the value of $\Neff$ that is suitable for our analysis. From \Aref{sec:Neff}, we estimate $\Neff$ to be between 
184 and 368, with $\Neff=184$ being the most pessimistic scenario and $\Neff=368$ being the most 
optimistic.}
    
Consider now the three scenarios described in \Sref{sec:ShearSys}, where KSB is used to correct for 
the PSF effects and the three levels of PSF modelling are assumed. For a hypothetical perfect PSF 
modelling technique, the shear errors in individual frames are already consistent with zero, so there is 
no need to discuss the combined results here. 
    
\begin{figure}  
   \begin{center}
   \includegraphics[height=2.4in]{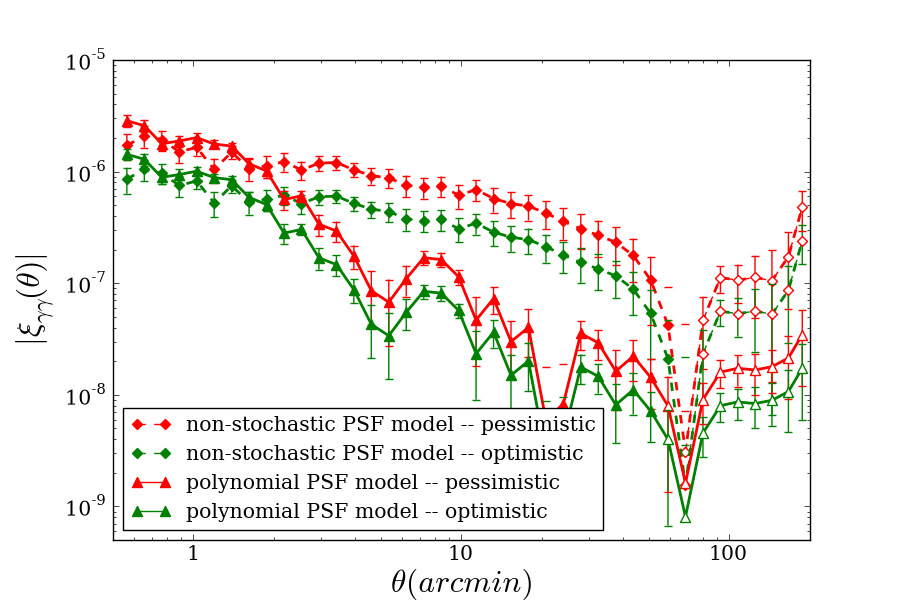}
   \caption{Absolute spurious shear correlation function after combining 10 years of $r$- and $i$-band 
   LSST data when a standard KSB algorithm is implemented and the PSF is modeled at different levels: 
   non-stochastic PSF knowledge only (dashed) and partial stochastic PSF knowledge from polynomial 
   interpolation of stars (solid). Red lines indicate the pessimistic case assuming $\Neff=184$ while the 
   green lines show the optimistic case when $\Neff=368$ is assumed. All curves are the medians of 20 
   different realisations under the fiducial observing condition, with the error bars indicating the standard 
   deviation in the 20 curves divided by $\sqrt{20}$. Negative values are plotted with open symbols.}
   \label{fig:combine_cf}  
\end{center}
\end{figure}  
   

For the second case, we know only the non-stochastic component of the PSF. In this case, spurious 
shear correlations result from not modelling any of the stochastic component of the PSF shape. In the 
combined dataset, the latter contribution can be estimated by taking the solid spurious shear correlation 
function in \Fref{fig:g_cf}(b) and multiply by $1/\Neff$ to account for the averaging of the stochastic spurious 
shear correlation. 

When both the non-stochastic and stochastic PSF components are modeled using a 5th-order 
polynomial model fitted to the stars, we assume that the smoothly varying non-stochastic PSF component 
is fully modeled and the spurious shear is mainly due to stochastic PSF modelling errors. The combined 
shear correlation function then can be estimated by scaling the spurious shear correlation function in 
\Fref{fig:g_cf}(c) by $1/\Neff$. 
    
The total expected spurious shear correlation functions from combining $\Neff$ exposures for the latter 
two cases are shown in \Fref{fig:combine_cf}.     

\subsection{Implication for constraints on cosmological parameters}
\label{sec:Cosmology}
We now interpret the spurious shear correlation function derived in 
\Sref{sec:Combine} in terms of the implied uncertainties on inferred cosmological parameters.

Since $\boldsymbol{\gamma}=0$ in all our analyses, we can identify the $\xi_{\gamma\gamma}$ 
measured in \Sref{sec:ShearSys} to be the ``additive spurious shear correlation function'' 
$\xi^{s}_{add}$ introduced in \citet{2006MNRAS.366..101H}. According to AR08, for several 
hypothetical forms of the spurious shear power spectrum, one can calculate the upper limits for allowed 
systematic errors of predictions of the major cosmological parameters via a simple extension to the Fisher 
Matrix formalism. This upper limits on the systematic errors are set so that the systematic errors do not exceed 
the statistical errors. In a survey with statistical power similar to the LSST survey, AR08 suggests the following 
limits on the spurious shear power spectrum: 
\begin{equation}
  \sigma^{2}_{sys}=\frac{1}{2\pi}\int_{\ell_{min}}^{\ell_{max}}  | C^{s}_{add}
  (\ell)|(\ell+1)d\ell\leq3\times 10^{-7} \;,
  \label{eq:sig_sys_limit}
\end{equation}
\noindent where $C^{s}_{add}$ is the power spectrum corresponding to $\xi^{s}_{add}$, which can be 
derived through \Eref{eq:ps2cf}.

\Eref{eq:sig_sys_limit} is in the form of shear power spectra, but our measurements are in the form 
of shear correlation functions. To properly connect our results to \Eref{eq:sig_sys_limit}, we revisit the 
hypothetical power spectrum used in AR08:
\begin{equation}
  C^{s}_{add}(\ell)=\frac{A_{0}}{\ell(\ell+1)}(n\: \log_{10}(\frac{\ell}{\ell_{0}})+1) \;,
  \label{eq:amara_ps}
\end{equation}
\noindent where $n$ is the slope of the log-linear power spectrum, $\ell_{0}$ is an arbitrary reference point 
chosen to be 700 in the paper and $A_{0}$ is the normalisation.	
 
Since the analytical form of \Eref{eq:amara_ps} is straightforward to integrate, we can use \Eref{eq:ps2cf} 
to find the correlation functions that correspond to power spectra in the form of \Eref{eq:amara_ps} for 
a range of $n$ and $A_{0}$ values. These correlation functions then can be compared to the spurious shear 
correlation functions in \Fref{fig:combine_cf} to determine the best matched $n$ and $A_{0}$ values. 
We calculate for this particular set of $n$ and $A_{0}$, the $ \sigma^{2}_{sys}$ values and compare with 
the target set via \Eref{eq:sig_sys_limit}. This process gives us an estimate of the level of uncertainties 
in the cosmological parameters when these forms of systematic errors in the shear correlation function 
are present. A more accurate estimate of $\sigma^{2}_{sys}$ can be obtained by the full Fisher Matrix calculation 
using these measured shear correlation functions. 

\begin{figure}  
   \begin{center}
   \includegraphics[height=2.4in]{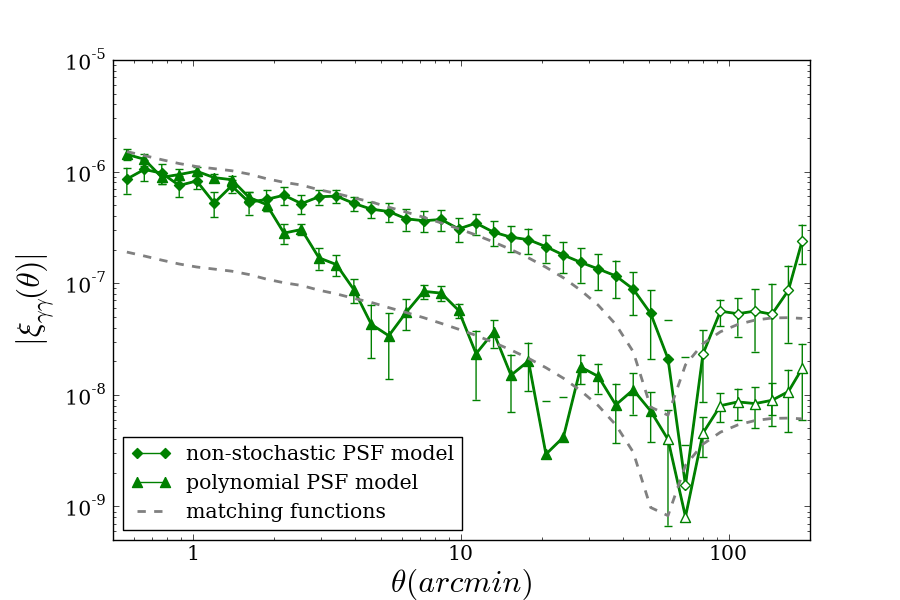}
   \caption{The two spurious shear correlation functions in the optimistic scenario (solid green) overlaid with 
   functional forms assumed by AR08 that visually match the level and approximate shapes of the spurious 
   shear correlation (dashed grey).  The green curves are the medians of 20 different realisations under the 
   fiducial observing condition, with the error bars indicating the standard deviation in the 20 curves divided 
   by $\sqrt{20}$. Negative values are plotted with open symbols.}
   \label{fig:g_fit}  
   \end{center}
 \end{figure}

We explore the parameter space $-3\leq n \leq 1$ and $10^{-9} \leq A_{0} \leq 10^{-5}$, which is chosen to be 
consistent with the ranges tested in AR08. In this range, we find that the family of functions is not always a good 
description for the spurious shear correlation function we measure from simulations. In particular, the sharp rising 
curve and the oscillations at small scales when polynomial PSF models are used cannot be properly modeled 
by the correlation function corresponding to the log-linear power spectra. As a result, we match scales only larger 
than $\sim3'$, knowing that in reality these smaller scales ( $<3'$) may not enter in constraining cosmology. 
The resulting $n$ and $A_{0}$ values as well as the corresponding $\sigma^{2}_{sys}$ values are listed in 
\Tref{table:sigma_sys}. \Fref{fig:g_fit} shows, for the optimistic case, the two spurious shear correlation functions 
overlaid by their functional-form counter parts. Note that these grey curves are not fits -- they are matched visually 
because the shapes of the measured shear correlation functions are quite different from the assumed functional 
forms.   

\begin{table}
  \centering
  \caption{For the 10-year combined $r$- and $i$-band data of LSST, the best matched $n$'s and $A_{0}$'s to the 
  spurious shear correlation function under different scenarios are listed. The numbers are measured with a KSB 
  pipeline and under two different PSF model assumptions. Two scenarios for effective number of exposures for 
  each field are assumed: the optimistic case corresponds to $\Neff=368$ and the pessimistic case corresponds to 
  $\Neff=184$.} 
  \begin{tabular}{c c c c c}
  \hline
  PSF model          & $\Neff$         & $n$      &      $\log_{10}(A_{0})$ & $\sigma^{2}_{sys}$ \\ \hline\hline
  \multirow{2}{*}{Non-stochastic}    & optimistic     &    \multirow{2}{*}{0.7}   &  $-5.7$ & 2.17$\times10^{-6}$\\
                                                             & pessimistic  &     &  $-5.4$ & 4.34$\times10^{-6}$\\ \hline  
  \multirow{2}{*}{Polynomial}          & optimistic     &    \multirow{2}{*}{0.7}   & $-6.6$  & 2.74$\times10^{-7}$\\
                                                             & pessimistic   &     & $-6.3$  & 5.46$\times10^{-7}$\\ \hline  
  \end{tabular}
  \label{table:sigma_sys}  
\end{table}

In \Tref{table:sigma_sys} we show that in the canonical weak lensing pipeline (KSB + polynomial PSF model), 
the spurious shear power spectrum we have measured from simulations is approximately 0.9 -- 1.8 times the 
statistical errors. Although the numbers imply that by using the current weak lensing pipeline, we are already 
reaching the level of systematics in shear measurements required for LSST, it should be understood thatÊnot 
all potential effects (such as shear calibration, galaxy modeling, photo-z estimation, 
chromatic PSF effect\footnote{The PSF shapes measured from the stars is different from the PSFs of the galaxies 
due to the differences between the SEDs of stars and galaxies.} \textit{etc.}) have yet been included. 
To ensure that the cosmic shear measurements from LSST is not systematics-limited 
\textit{after} considering all the other systematic errors, we will need shear measurement methods more 
sophisticated than what we used in this study. 

We can trace the source of these systematic errors to improper modelling of the stochastic PSF using polynomial 
functions, which needs to be reduced when developing the next generation of shear measurement algorithms. 
On the other hand, if only the non-stochastic errors are modeled, the spurious shear due to not modelling the 
stochastic PSF results in a $\sigma^{2}_{sys}$ value one order of magnitude greater than the target value. This 
implies that the stochastic PSF components do not average out enough by themselves if not corrected, even when 
the full dataset is combined. We thus show the importance of modelling the stochastic as well as the 
non-stochastic components of the PSF. 

\chihway{
We have shown here that given a typical weak lensing pipeline, the major physical effects in an LSST 
observation will not seriously limit LSST, provided that the number of exposures in the combined dataset and 
the image quality are as expected.}

\subsection{Effect of simplifications}
\label{sec:simplifications}
\chihway{At this point we summarize the major assumptions that underlie our analysis to provide context for our results: }

\chihway{
First, we have deliberately designed our simulations and the analysis we performed to minimize algorithm-dependent 
contributions to the errors. In particular, we used circular Gaussians as our galaxy models, invoked KSB as our shear 
measurement method, and performed an artificial ``perfect calibration'' for the KSB pipeline. Thus the results derived in 
\srefa{sec:Combine} and \srefb{sec:Cosmology} only take account of the algorithm-independent part of the additive 
spurious shear correlation function. In particular, recent work \citep{2003MNRAS.343..459H, 2012arXiv1203.5050R, 
2012MNRAS.tmp.3383M} has shown that the algorithm-dependent shear errors are strongly affected by noise (the so 
called ``noise bias''), which arises from using very low SNR galaxies. In our analyses, this factor is suppressed through 
the use of a simplistic galaxy model. However, given the low SNR ($\sim$8) of our fiducial galaxy, the noise bias for 
realistic weak lensing galaxies may not be negligible. }
 
\chihway{ 
Second, we have not taken into account more sophisticated schemes for combining shear 
measurements from multiple exposures. As suggested by \citet{2006JCAP...02..001J}, the stochastic component in the 
shear errors can be eliminated by dividing the full dataset into sub-groups of exposures and only correlating shear 
measurements between different sub-groups. We provide a brief discussion in \Aref{sec:cross_correlation} of such 
implementations, but have not investigated the full power of these alternative approaches in this paper.}

\chihway{
Third, as mentioned in \Sref{sec:Introduction}, this work is based on the projected two-point shear correlation function. 
In a full weak lensing analyses where lensing tomography and higher-order statistics are used, additional constraints 
may arise. However, the combination of all these different statistics may also be useful in mitigating certain systematic 
effects.}

Finally, we also note that the applicability of the extended Fisher Matrix formulae in AR08 and therefore 
\Eref{eq:sig_sys_limit} to our analyses depends on some specific assumptions regarding the statistical properties of 
the spurious shear contributions we measure in the simulations. The main assumption in AR08 is that the higher-order 
statistical properties of the spurious shear are similar to those of the true shear -- a Gaussian random field -- so that the 
covariance matrix for the spurious shear power spectrum is close to diagonal. (This is implicitly assumed when deriving 
Equation 10 from Equation 9 in AR08.) Since the statistical properties of the spurious shear depend on its physical origin, 
they are not guaranteed to be Gaussian. However, in our simulations, the effect of such non-Gaussian spurious shear is 
likely to be small compared to the Gaussian component generated from counting statistics and the various stochastic 
effects; therefore the results derived in \Sref{sec:Cosmology} should be sufficiently robust.

\section{Conclusions}
\label{sec:Conclusion}

In this paper, we have carried out a bottom-up, quantitative study of the potential systematic errors in cosmic shear 
measurements for future LSST-like surveys using high fidelity simulations. 

Simulations are generated using \phosim, a photon-by-photon Monte Carlo ray-tracing software that models all 
major physical effects from the top of the atmosphere down through the detectors. Specifically, we have generated 
a suite of special simulations in order to isolate the systematic errors in ellipticity and shear measurements caused 
by different physical effects, which would have been impossible to achieve with a real telescope. 

We identify the most important physical effects in terms of their impact on ellipticity measurements and classify them 
into two classes: non-stochastic and stochastic. The ellipticity errors and their correlation properties caused by each 
individual effect are then quantified in a systematic way. We find that, in a single LSST exposure:
\begin{itemize} 
  \item Ellipticity errors due to counting statistics dominate the total ellipticity errors, whereas ellipticity errors due to 
  atmospheric and instrumental effects dominate the total ellipticity error correlation function. 
  \item The ellipticity error correlation function due to non-stochastic effects is one order of magnitude smaller than 
  that due to stochastic effects.
\end{itemize}  

For shear measurement, we identify three steps in a canonical weak lensing pipeline that lead to spurious shear, two 
of which are dominated by the specific algorithm chosen for PSF characterisation and deconvolution, which we have 
not investigated in detail. The third step involves modelling the PSF spatial variation with scattered stars. We carry out 
the full analyses with a standard weak lensing algorithm and quantify the spurious shear correlation under different 
assumptions about the PSF model. We draw several conclusions:
\begin{itemize} 
  \item With perfect PSF knowledge, systematics induced by the algorithm in an idealised KSB implementation are   
  negligible.
  \item Not correcting for the stochastic component of the PSF shape introduces large shear systematics in the 
  correlation function.
  \item A conventional PSF modelling scheme using polynomial interpolation of stars can partially model the 
  stochastic PSF contribution, but the inflexibility of the functional form of polynomials limits the power of this method. 
\end{itemize}  
  
The single-exposure results are then extrapolated to the full combined 10-year dataset, and finally interpreted in 
terms of the constraints on dark energy parameters according to an extended Fisher Matrix calculation from AR08. 
We draw several conclusions:

\begin{itemize} 
  \item By using a canonical weak lensing analysis pipeline, the systematic errors in the spurious shear correlation 
  function induced by \chihway{the major physical effects}, after combining the 10 years of LSST data, is at a level 
  approaching the statistical errors. 
  \item The errors mainly come from imperfect modelling of the stochastic PSF, which has not been studied in 
  detail in the past. This calls for better basis functions that can characterise structures in the stochastic PSF 
  variation on all scales. 
\end{itemize}  

Finally, this analysis is done under several assumptions and simplifications, which may need to be taken into account 
when interpreting the results:

\begin{itemize} 
  \item \chihway{We have designed the simulations and analysis to avoid algorithm-dependence of this analysis. Algorithm 
  errors will need to be estimated and combined with the results here to yield the total shear systematic errors.} 
 \item \chihway{A simple scheme is used for combination of shear measurements in multiple exposures. More intelligent 
 use of the data can potentially give better results.}
  \item Only a projected 2D two-point correlation function is analysed. By implementing weak lensing tomography or 
  higher-order statistics, some of the spurious shear can be mitigated. 
  \item We adopt the results from AR08 to interpret the spurious shear correlation function in terms of its effect on 
  the uncertainty in predicting cosmological parameters, which implicitly assume that the spurious shear has statistical 
  properties similar to the true shear.  
\end{itemize}  

\section*{Acknowledgments}
LSST project activities are supported in part by the National Science
Foundation through Governing Cooperative Agreement 0809409 managed by the
Association of Universities for Research in Astronomy (AURA), and the
Department of Energy under contract DE-AC02-76-SFO0515 with the SLAC
National Accelerator Laboratory. Additional LSST funding comes from private
donations, grants to universities, and in-kind support from LSSTC Institutional
Members.

We thank Bhuvnesh Jain for many useful discussions in the progress of writing this paper. We 
thank Patricia Burchat and Seth Digel and the anonymous referee for useful comments which 
have helped improve this paper substantially.

\label{lastpage}


\begin{thebibliography}{}

\bibitem[\protect\citeauthoryear{{Albrecht} et~al.,}{{Albrecht}
  et~al.}{2006}]{2006APS..APR.G1002A}
{Albrecht} A.,  et~al., 2006, APS April Meeting Abstracts, pp G1002+

\bibitem[\protect\citeauthoryear{{Amara} \& {R{\'e}fr{\'e}gier}}{{Amara} \&
  {R{\'e}fr{\'e}gier}}{2007}]{2007MNRAS.381.1018A}
{Amara} A.,  {R{\'e}fr{\'e}gier} A.,  2007, \mnras, 381, 1018

\bibitem[\protect\citeauthoryear{{Amara} \& {R{\'e}fr{\'e}gier}}{{Amara} \&
  {R{\'e}fr{\'e}gier}}{2008}]{2008MNRAS.391..228A}
{Amara} A.,  {R{\'e}fr{\'e}gier} A.,  2008, \mnras, 391, 228

\bibitem[\protect\citeauthoryear{{Bacon}, {R\'efr\'egier} \& {Ellis}}{{Bacon}
  et~al.}{2000}]{2000MNRAS.318..625B}
{Bacon} D.~J.,  {R\'efr\'egier} A.~R.,    {Ellis} R.~S.,  2000, \mnras, 318,
  625

\bibitem[\protect\citeauthoryear{{Bartelmann} \& {Schneider}}{{Bartelmann} \&
  {Schneider}}{2001}]{2001PhR...340..291B}
{Bartelmann} M.,  {Schneider} P.,  2001, \physrep, 340, 291

\bibitem[\protect\citeauthoryear{{Benjamin} et~al.,}{{Benjamin}
  et~al.}{2007}]{2007MNRAS.381..702B}
{Benjamin} J.,  et~al., 2007, \mnras, 381, 702

\bibitem[\protect\citeauthoryear{{Bertin} \& {Arnouts}}{{Bertin} \&
  {Arnouts}}{1996}]{1996A&AS..117..393B}
{Bertin} E.,  {Arnouts} S.,  1996, \aaps, 117, 393

\bibitem[\protect\citeauthoryear{{Bridle} et~al.,}{{Bridle}
  et~al.}{2010}]{2010MNRAS.405.2044B}
{Bridle} S.,  et~al., 2010, \mnras, 405, 2044

\bibitem[\protect\citeauthoryear{{Chang} et~al.,}{{Chang}  et~al.}{2012}]{C12}
{Chang} C.,  et~al., 2012, in preparation

\bibitem[\protect\citeauthoryear{{Connolly} et~al.,}{{Connolly}
  et~al.}{2010}]{2010SPIE.7738E..53C}
{Connolly} A.~J.,  et~al., 2010, in SPIE Conference Series Vol.~7738 of SPIE
  Conference Series, {Simulating the LSST system}

\bibitem[\protect\citeauthoryear{{De Vries} et~al.,}{{De Vries}
  et~al.}{2007}]{2007ApJ...662..744D}
{De Vries} W.~H.,  et~al., 2007, \apj, 662, 744

\bibitem[\protect\citeauthoryear{{Hetterscheidt} et~al.,}{{Hetterscheidt}
  et~al.}{2007}]{2007A&A...468..859H}
{Hetterscheidt} M.,  et~al., 2007, \aap, 468, 859

\bibitem[\protect\citeauthoryear{{Heymans} et~al.,}{{Heymans}
  et~al.}{2006}]{2006MNRAS.368.1323H}
{Heymans} C.,  et~al., 2006, \mnras, 368, 1323

\bibitem[\protect\citeauthoryear{{Heymans} et~al.,}{{Heymans}
  et~al.}{2012}]{2012MNRAS.421..381H}
{Heymans} C.,  et~al., 2012, \mnras, 421, 381

\bibitem[\protect\citeauthoryear{{Hirata} \& {Seljak}}{{Hirata} \&
  {Seljak}}{2003}]{2003MNRAS.343..459H}
{Hirata} C.,  {Seljak} U.,  2003, \mnras, 343, 459

\bibitem[\protect\citeauthoryear{{Hoekstra} et~al.,}{{Hoekstra}
  et~al.}{1998}]{1998ApJ...504..636H}
{Hoekstra} H.,  et~al., 1998, \apj, 504, 636

\bibitem[\protect\citeauthoryear{{Hoekstra}, {Mellier}, {van Waerbeke},
  {Semboloni}, {Fu}, {Hudson}, {Parker}, {Tereno} \& {Benabed}}{{Hoekstra}
  et~al.}{2006}]{2006ApJ...647..116H}
{Hoekstra} H.,  {Mellier} Y.,  {van Waerbeke} L.,  {Semboloni} E.,  {Fu} L.,
  {Hudson} M.~J.,  {Parker} L.~C.,  {Tereno} I.,    {Benabed} K.,  2006, \apj,
  647, 116

\bibitem[\protect\citeauthoryear{{Hu}}{{Hu}}{1999}]{1999ApJ...522L..21H}
{Hu} W.,  1999, \apjl, 522, L21

\bibitem[\protect\citeauthoryear{{Hu} \& {Tegmark}}{{Hu} \&
  {Tegmark}}{1999}]{1999ApJ...514L..65H}
{Hu} W.,  {Tegmark} M.,  1999, \apjl, 514, L65

\bibitem[\protect\citeauthoryear{{Huff} et~al.,}{{Huff}
  et~al.}{2011}]{2011arXiv1112.3143H}
{Huff} E.~M.,  et~al., 2011, ArXiv e-prints

\bibitem[\protect\citeauthoryear{{Huterer}, {Takada}, {Bernstein} \&
  {Jain}}{{Huterer} et~al.}{2006}]{2006MNRAS.366..101H}
{Huterer} D.,  {Takada} M.,  {Bernstein} G.,    {Jain} B.,  2006, \mnras, 366,
  101

\bibitem[\protect\citeauthoryear{{Ivezic} et~al.,}{{Ivezic}
  et~al.}{2008}]{2008arXiv0805.2366I}
{Ivezic} Z.,  et~al., 2008, ArXiv e-prints: astro-ph/0805.2366

\bibitem[\protect\citeauthoryear{{Ivezic} et~al.,}{{Ivezic}
  et~al.}{2011}]{SRD}
{Ivezic} Z.,  et~al., 2011, The LSST System Science Requirement Document v5.2.3

\bibitem[\protect\citeauthoryear{{Jain}, {Jarvis} \& {Bernstein}}{{Jain}
  et~al.}{2006}]{2006JCAP...02..001J}
{Jain} B.,  {Jarvis} M.,    {Bernstein} G.,  2006, Journal of Cosmology and
  Astroparticle Physics, 2, 1

\bibitem[\protect\citeauthoryear{{Jain} \& {Seljak}}{{Jain} \&
  {Seljak}}{1997}]{1997ApJ...484..560J}
{Jain} B.,  {Seljak} U.,  1997, \apj, 484, 560

\bibitem[\protect\citeauthoryear{{Jarvis} \& {Jain}}{{Jarvis} \&
  {Jain}}{2004}]{2004astro.ph.12234J}
{Jarvis} M.,  {Jain} B.,  2004, ArXiv e-prints: astro-ph/0412234

\bibitem[\protect\citeauthoryear{{Jarvis}, {Schechter} \& {Jain}}{{Jarvis}
  et~al.}{2008}]{2008arXiv0810.0027J}
{Jarvis} M.,  {Schechter} P.,    {Jain} B.,  2008, ArXiv e-prints:
  astro-ph/0810.0027

\bibitem[\protect\citeauthoryear{{Jee} \& {Tyson}}{{Jee} \&
  {Tyson}}{2011}]{2011PASP..123..596J}
{Jee} M.~J.,  {Tyson} J.~A.,  2011, \pasp, 123, 596

\bibitem[\protect\citeauthoryear{{Kaiser}}{{Kaiser}}{1998}]{1998ApJ...498...26%
K}
{Kaiser} N.,  1998, \apj, 498, 26

\bibitem[\protect\citeauthoryear{{Kaiser}, {Squires} \& {Broadhurst}}{{Kaiser}
  et~al.}{1995}]{1995ApJ...449..460K}
{Kaiser} N.,  {Squires} G.,    {Broadhurst} T.,  1995, \apj, 449, 460

\bibitem[\protect\citeauthoryear{{Kaiser}, {Wilson} \& {Luppino}}{{Kaiser}
  et~al.}{2000}]{2000astro.ph..3338K}
{Kaiser} N.,  {Wilson} G.,    {Luppino} G.~A.,  2000, ArXiv e-prints:
  astro-ph/0003338

\bibitem[\protect\citeauthoryear{{Kitching} et~al.,}{{Kitching}
  et~al.}{2012a}]{2012MNRAS.423.3163K}
{Kitching} T.~D.,  et~al., 2012a, \mnras, 423, 3163

\bibitem[\protect\citeauthoryear{{Kitching} et~al.,}{{Kitching}
  et~al.}{2012b}]{2012arXiv1204.4096K}
{Kitching} T.~D.,  et~al., 2012b, ArXiv e-prints: astro-ph/1204.4096

\bibitem[\protect\citeauthoryear{Kolmogorov}{Kolmogorov}{1992}]{Kolmogorov1941}
Kolmogorov A.,  1992, Dokl. Akad. Nauk SSSR, 30, 301

\bibitem[\protect\citeauthoryear{{Krabbendam} et~al.,}{{Krabbendam}
  et~al.}{2010}]{2010AAS...21540105K}
{Krabbendam} V.,  et~al., 2010, in AAS Meeting Abstracts \#215 Vol.~42 of
  Bulletin of the AAS, {LSST Operations Simulator}.
pp 401.05--+

\bibitem[\protect\citeauthoryear{{Lin} et~al.,}{{Lin}
  et~al.}{2011}]{2011arXiv1111.6622L}
{Lin} H.,  et~al., 2011, ArXiv e-prints

\bibitem[\protect\citeauthoryear{{Luppino} \& {Kaiser}}{{Luppino} \&
  {Kaiser}}{1997}]{1997ApJ...475...20L}
{Luppino} G.~A.,  {Kaiser} N.,  1997, \apj, 475, 20

\bibitem[\protect\citeauthoryear{{Massey} et~al.,}{{Massey}
  et~al.}{2007}]{2007MNRAS.376...13M}
{Massey} R.,  et~al., 2007, \mnras, 376, 13

\bibitem[\protect\citeauthoryear{{Melchior} \& {Viola}}{{Melchior} \&
  {Viola}}{2012}]{2012MNRAS.tmp.3383M}
{Melchior} P.,  {Viola} M.,  2012, \mnras, p.~3383

\bibitem[\protect\citeauthoryear{{Paulin-Henriksson}, {Amara}, {Voigt},
  {R\'efr\'egier} \& {Bridle}}{{Paulin-Henriksson}
  et~al.}{2008}]{2008A&A...484...67P}
{Paulin-Henriksson} S.,  {Amara} A.,  {Voigt} L.,  {R\'efr\'egier} A.,
  {Bridle} S.~L.,  2008, \aap, 484, 67

\bibitem[\protect\citeauthoryear{{Peterson} et~al.,}{{Peterson}
  et~al.}{2009}]{P10}
{Peterson} J.~R.,  et~al., 2009, {LSST Science Book, Version 2.0, Chapter 3.3}

\bibitem[\protect\citeauthoryear{{Peterson} et~al.,}{{Peterson}
  et~al.}{2012}]{P12}
{Peterson} J.~R.,  et~al., 2012, in preparation

\bibitem[\protect\citeauthoryear{{Poyneer}, {van Dam} \& {V{\'e}ran}}{{Poyneer}
  et~al.}{2009}]{2009JOSAA..26..833P}
{Poyneer} L.,  {van Dam} M.,    {V{\'e}ran} J.-P.,  2009, Journal of the
  Optical Society of America A, 26, 833

\bibitem[\protect\citeauthoryear{{R{\'e}fr{\'e}gier}
  et~al.,}{{R{\'e}fr{\'e}gier}  et~al.}{2012}]{2012arXiv1203.5050R}
{R{\'e}fr{\'e}gier} A.,  et~al., 2012, ArXiv e-prints

\bibitem[\protect\citeauthoryear{{Schneider}, {Kilbinger} \&
  {Lombardi}}{{Schneider} et~al.}{2005}]{2005A&A...431....9S}
{Schneider} P.,  {Kilbinger} M.,    {Lombardi} M.,  2005, \aap, 431, 9

\bibitem[\protect\citeauthoryear{{Schneider} \& {Lombardi}}{{Schneider} \&
  {Lombardi}}{2003}]{2003A&A...397..809S}
{Schneider} P.,  {Lombardi} M.,  2003, \aap, 397, 809

\bibitem[\protect\citeauthoryear{{Schrabback} et~al.,}{{Schrabback}
  et~al.}{2010}]{2010A&A...516A..63S}
{Schrabback} T.,  et~al., 2010, \aap, 516, A63+

\bibitem[\protect\citeauthoryear{{Semboloni} et~al.,}{{Semboloni}
  et~al.}{2006}]{2006A&A...452...51S}
{Semboloni} E.,  et~al., 2006, \aap, 452, 51

\bibitem[\protect\citeauthoryear{{Taylor}}{{Taylor}}{1938}]{1938RSPSA.164..476%
T}
{Taylor} G.~I.,  1938, Royal Society of London Proceedings Series A, 164, 476

\bibitem[\protect\citeauthoryear{{Tyson}}{{Tyson}}{2002}]{2002SPIE.4836...10T}
{Tyson} J.~A.,  2002, in {J.~A.~Tyson \& S.~Wolff} ed., SPIE Conference Series
  Vol.~4836 of SPIE Conference Series, {Large Synoptic Survey Telescope:
  Overview}.
pp 10--20

\bibitem[\protect\citeauthoryear{{Wittman}}{{Wittman}}{2005}]{2005ApJ...632L..%
.5W}
{Wittman} D.,  2005, \apjl, 632, L5

\bibitem[\protect\citeauthoryear{{Wittman} et~al.,}{{Wittman}
  et~al.}{2000}]{2000Natur.405..143W}
{Wittman} D.~M.,  et~al., 2000, \nat, 405, 143

\bibitem[\protect\citeauthoryear{{Zhang}}{{Zhang}}{2010}]{2010MNRAS.403..673Z}
{Zhang} J.,  2010, \mnras, 403, 673

\end{thebibliography}

\bsp


\appendix

\section{Physical models in \Phosim}
\label{sec:ImSim_model_app}

\subsection{Optics and optics perturbations}
\label{sec:ImSim_model_optics}

\phosim builds in the most up-to-date optics design of the instrument. This includes detailed 
specifications of the dimensions and wavelength response of each optical element from the 
engineering design (the three mirrors, three lenses and the filter), characteristics of the 
backside-illuminated thick CCD detectors, and other telescope components such as the 
shutter and spider, scattered light and tracking mechanisms. The \phosim version used in 
this paper is based upon the optics baseline design version 3.3. 
        
In addition to the design, \phosim also models the level of residual wavefront errors after a 
typical correction from the AOS has been made. The effects of these residuals are modeled using 
hundreds of parameters that displace or deform the different optical components, causing the PSF 
to degrade from the design within levels allowed by the engineering requirements on the AOS. 
This approach is 
different from an exact simulation of the AOS, which would have to take into account
the history of the wavefront measurements over a period of hours as the survey proceeds. The 
latter approach would require a vast increase in the number of exposure simulations that are 
performed, and is thus computationally impractical. Therefore, the current \phosim takes the alternative 
approach of modelling residuals of the full control loop instead. Since exposures on the same patch 
of sky are usually separated by a few days, which is much longer than a typical time scale for the AOS  
updates, the assumptions involved in this procedure are well justified.

We further classify the residual wavefront errors, or optics errors in \phosim, into two classes. 
The first class accounts for errors that originate from fabrication and integration, or are 
introduced by gravity and the thermal environment of the telescope. The former are permanent functions 
of the response, while the latter are by nature highly repeatable, and the zeroth-order corrections will be 
implemented by the AOS according to a pre-calibrated look-up table. However, the AOS cannot 
perfectly compensate for all distortions given the limited number of degrees of freedom that are 
actively controlled; therefore, there will be repeatable optics errors, which are ``non-stochastic''. The 
characteristic amplitudes of these perturbations used in \phosim are derived from finite element modelling 
of the telescope and camera under appropriate gravity and thermal loads. The second class includes the 
residual wavefront errors as well as actuator errors and wind shake. Since these effects are random in 
nature and do not show repeated patterns across exposures, they can be measured only by monitoring 
the wavefront errors in real time. The imperfect correction by the AOS for these effects introduces 
``stochastic'' optics errors which are uncorrelated from exposure to exposure. For the purpose of this paper, 
we deliberately implement the optics model in \phosim in a way that allows the two classes of optics errors 
to be separated -- this is because we are interested in how they separately enter into the shear 
measurements, as discussed in more detail in \Sref{sec:Framework_EllipSys}. 

The typical levels of the two classes of optics errors on the major optical elements used in this analysis are 
listed in \Tref{table:optics_parameters}. Note that the numbers corresponding to this classification scheme 
are based on the current engineering specifications \citep[][and internal documents]{SRD}, but the results are 
easily scalable if different specifications are eventually adopted..

\begin{table*}
\begin{center}
  \caption{Specifications of the major optics errors modeled in \phosim. We describe the surface height 
  variations on the mirrors using 2nd- through 5th- order Zernike polynomials, where each polynomial is 
  normalised individually with some amplitude according to typical values observed in existing systems. We 
  list here only the range for the four amplitudes used.}  
  \label{table:optics_parameters}
  \begin{tabular}{c c c  c c }
  \hline
  Source of error                    & Stochastic or not                & Random sampling             & Physical parameter      &  RMS values  \\ \hline \hline
                                                 & \multirow{2}{*}{non-stochastic}   & \multirow{2}{*}{Gaussian} & misalignment & $1.08 \times 10^{-2}$ (mm)\\ 
  Solid-body optics errors    &                                      &                                                & tip/tilt                &  $2.34 \times 10^{-6}$ (rad)\\
      in the mirrors                    & \multirow{2}{*}{stochastic} &    \multirow{2}{*}{Gaussian}       & misalignment & $1.89 \times 10^{-2}$ (mm)\\ 
                                                 &                                      &                                                & tip/tilt                 & $4.09 \times 10^{-6}$ (rad)\\ \hline
  Surface height variation    & non-stochastic           & \multirow{2}{*}{Gaussian}  & coefficients of Zernike   & (0.38 -- 2.20) $ \times 10^{-4}$ (mm)\\ 
  in the mirrors                        &stochastic                    &                                                &  polynomial expansion & (0.76 -- 4.40) $ \times 10^{-4}$ (mm)\\ \hline
                                                 & \multirow{2}{*}{non-stochastic}    &  \multirow{2}{*}{Gaussian} & misalignment  & $3.24 \times 10^{-2}$ (mm)\\ 
  Solid-body optics errors    &                                       &                                                 &tip/tilt                 & $1.58 \times 10^{-4}$  (rad)\\ 
  in the camera                       & \multirow{2}{*}{stochastic}  &     \multirow{2}{*}{Gaussian}    & misalignment  & $5.67 \times 10^{-2}$ (mm)\\ 
                                                 &                                       &                                                 & tip/tilt                & $3.77 \times 10^{-4}$  (rad)\\ \hline
  Sensor surface displacement      & \multirow{2}{*}{non-stochastic}    &  Zernike expansion based       &  height    & \multirow{2}{*}{5$\times10^{-3}$ (mm)}\\ 
  from ideal focal plane         &                                      & on laboratory measurements & variation &      \\ \hline
  \end{tabular}
\end{center}
\end{table*}  


\subsection{Tracking errors}
\label{sec:ImSim_model_tracking}

Tracking errors in \phosim are modeled by a Gaussian random walk of the telescope pointing 
in each of the three directions: azimuth, elevation and rotation, where a step is taken every 0.1 
second throughout the exposure. The effect of the tracking error integrated over 15 seconds is 
to yield a root-mean-squared (RMS) error of $\sim0.02$" in the azimuth and elevation directions, 
and  $\sim1$" in rotation. 

\subsection{The atmospheric model}
\label{sec:ImSim_model_atm}
Due to the short exposure time of LSST, atmospheric effects on image distortions become 
more pronounced in a single exposure than what is usually seen in longer exposures. Since 
these atmosphere-induced distortions arise from turbulent structures in the air density, we 
can also expect spatial structures in the image shape distortions across the field which are 
associated with the turbulent structures. \citet{2007ApJ...662..744D} first showed via simulations 
that the effect of the atmospheric turbulence on PSF shape distortion averages out rapidly with 
exposure time, while \citet{2005ApJ...632L...5W} and \citet{2012MNRAS.421..381H} measured 
the correlation of PSF shape distortions in short exposure data and discuss their potential effect on 
weak lensing. 
        
Similar to the pure atmospheric simulations in \citet{2007ApJ...662..744D}, the atmosphere in 
\phosim is modeled by multiple layers of moving atmospheric screens. The ``frozen screen 
approximation'' is justified since the time scale for the shapes of turbulent cells to change 
significantly is much longer than the time required for turbulence cells to pass through the field of 
view, given the typical wind speeds of a few meters per second \citep{1938RSPSA.164..476T, 
2009JOSAA..26..833P}. 
        
The heart of the atmospheric model is a set of multi-scale, multi-layer frozen Kolmogorov screens 
\citep{Kolmogorov1941}. These atmospheric screens are constructed according to a full 3D 
Kolmogorov spectrum with assigned parameters including the structure function, inner scale, 
outer scale, wind speed and wind direction. All parameters are modeled from existing atmospheric 
data taken near the LSST site. In \citet{P12}, we explain the theoretical justification of 
the approach we have taken, as well as the major innovations in the \phosim atmospheric model.

\section{KSB formulas}
\label{sec:KSB}

The following formulas are the foundation for performing a weak lensing measurement using the KSB 
algorithm. We use the Einstein summation convention.
 
First, stars and galaxies are measured with the ``getshape'' \imcat routine, which assigns to each object 
the following shape parameters: the complex ellipticity ($\varepsilon_{\alpha}$), the smear polarisability 
($P^{sm}_{\alpha\beta}$) and the shear polarisability ($P^{sh}_{\alpha\beta}$), where $\alpha,\beta=1,2$. 
$\varepsilon_{\alpha}$ is defined in \erefa{eq:ellipticity} and \erefb{eq:moments}, while 
$P^{sm}_{\alpha\beta}$ and $P^{sh}_{\alpha\beta}$ are calculated through
\begin{equation}
  P_{\alpha\beta}^{sm}=X_{\alpha\beta}^{sm}-\varepsilon_{\alpha}
  \varepsilon_{\beta}^{sm}\;,
\end{equation}
\begin{equation}
  P_{\alpha\beta}^{sh}=X_{\alpha\beta}^{sh}-\varepsilon_{\alpha}
  \varepsilon_{\beta}^{sh}\;.
\end{equation}
\noindent $X_{\alpha\beta}^{sm}$, $X_{\alpha\beta}^{sh}$, $\varepsilon_{\beta}^{sm}$ 
and $\varepsilon_{\beta}^{sh}$ are derived through combinations of weighted second moments of the light 
profile of the object $f(x_{1},x_{2})$ and derivatives of the weighting function $W(x_{1},x_{2})$ \citep[see][ for 
definitions of these quantities]{1995ApJ...449..460K}. Each component $\varepsilon_{\alpha}$, 
$P^{sm}_{\alpha\beta}$ and $P^{sh}_{\alpha\beta}$ for the stars is interpolated to the galaxys' locations and 
recalculated with the galaxys' weighting functions. The remeasured quantities  will be used to construct the 
PSF model for the galaxy.

The anisotropic PSF effects on the galaxy's ellipticity are first corrected through 
\begin{equation}
  \delta \varepsilon_{\alpha}=P_{\alpha\beta}^{sm}p_{\beta}\;,
\end{equation}
\noindent where 
\begin{equation}
  p_{\alpha}=(P^{*,sm})^{-1}_{\alpha \beta}\varepsilon^{*}_{\beta}\;.
\end{equation}
\noindent The superscript ``$*$'' indicates parameters of the PSF model. Shear $g_{\alpha}$ changes 
the galaxy's ellipticity by
\begin{equation}
\delta \varepsilon_{\alpha}=P_{\alpha\beta}^{sh}g_{\beta}\;.
\end{equation}
 
Finally, we need to correct for the weighting and circular seeing effects to get the final shear estimate for 
each galaxy by replacing $P_{\alpha\beta}^{sh}$ with the shear susceptibility $P_{\alpha\beta}^{\gamma}$, 
where
\begin{equation}
  P_{\alpha\beta}^{\gamma}=P^{sh}_{\alpha \beta}-P^{sm}_{\alpha \gamma}
  (P^{*,sm})^{-1}_{\gamma \delta}P^{*,sh}_{\delta \beta}\;.
\end{equation}
$P_{\alpha\beta}^{\gamma}$ is eventually replaced by $\frac{1}{2}tr[P^{\gamma}]$ in our approach, 
similar to the ``ES2'' method in \citet{2007MNRAS.376...13M}.

\section{$1/N$ scaling for the stochastic spurious shear correlation}
\label{sec:Nscaling}
\chihway{
To demonstrate the $1/N$ scaling for the stochastic spurious shear correlation, we use the 20 simulated 
shear catalogues used in \Sref{sec:ShearSys_star_psf}, where we have argued that most of the shear 
errors in these catalogues are stochastic. We then consider averaging shear measurements of the 
same galaxy in the first $N$ different frames, and calculate the shear correlation function for the averaged   
shear catalogue as a function of $N$.}

\begin{figure}  
   \begin{center}
   \includegraphics[height=2.4in]{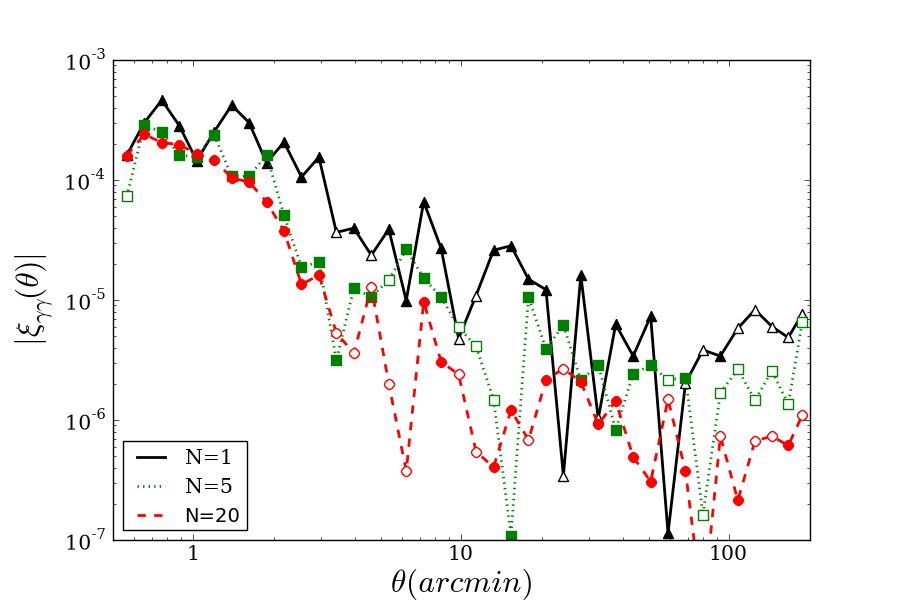}
   \caption{\chihway{The spurious shear correlation function for the mean shear catalogue for $N$ exposures, where 
   $N=1$ (solid black), 5 (dotted green) and 20 (dashed red). The shear catalogue is obtained via standard KSB 
   algorithm and 5th-order polynomial PSF models. Negative values are plotted with open symbols.}}
   \label{fig:ave_cf}  
   \end{center}
 \end{figure}

\chihway{
The results for $N=1,5,20$ are shown in \Fref{fig:ave_cf}. We observe that, although the low statistics causes 
the data to be quite noisy, the spurious shear correlation function does roughly follow the $1/N$ scaling. This 
supports our argument in \Sref{sec:Combine}, where we extrapolate our single-exposure measurements 
to the full 10-year LSST dataset. }

\section{Effective number of exposures}
\label{sec:Neff}
    
In the 10-year period of observation planned for LSST, every patch of sky is imaged approximately 
386 times, each with 15-second exposures (the number doubles to 772 if $r$ and $i$ bands 
are combined); however, because of variation in the observing conditions and the galaxy properties 
themselves, not all galaxies have good shape measurements in all exposures. The power of combining a 
multi-epoch dataset is that one has the freedom to weight the contributions of a single galaxy shape 
measurement in each of the different exposures according to the image quality, and thus 
extract the maximum information from the noisy dataset. This effectively means that the ``stochastic'' 
spurious shear correlation will not cancel as fast as $1/772$ under this framework, because some of 
the exposures will be down-weighted due to their poor image quality. Instead, the spurious shear correlation 
will decrease only as fast as $1/\Neff$, where $\Neff$ is the ``effective number of exposures'' for the entire 
dataset. $\Neff$ depends on the detailed distribution of the observing conditions, the galaxy distributions 
(brightness, size, redshift), as well as the analysis pipeline used. The precise determination of $\Neff$ is 
complicated and beyond the scope of this paper; however, we can provide optimistic and pessimistic estimates 
for $\Neff$ to bound the final results in a reasonable range.  
    
In this work, we have chosen the fiducial observing conditions (\Tref{table:fiducial_params}) to correspond 
to roughly the median condition of the best 50\% of the dataset in terms of image quality. This gives 
approximately 184 exposures in the $r$ band as well as the $i$ band, or $\sim$368 exposures for the entire 
weak lensing dataset for each field. We view this as an \textit{optimistic} estimation, since it has been shown 
\citep{2003MNRAS.343..459H, 2012arXiv1203.5050R, 2012MNRAS.tmp.3383M} that for PSF sizes 
approaching the size of average galaxies (for part of the 368 exposures), the systematics can grow nonlinearly 
and the median spurious shear of these 368 exposures can be larger than that shown in \Fref{fig:g_cf}. 
\chihway{Furthermore, as mentioned in \Sref{sec:s_ns}, errors in the 2 exposures in the same visit may 
be correlated, effectively lowering $\Neff$ for the full dataset.} A more 
conservative estimation is to use only the best 25\% of the exposures to make cosmic shear measurements. In 
this case we have a total of 184 exposures and the results are likely to be pessimistic, since the median 
spurious shear correlation of these 184 exposures is likely to be better than those measured in \Fref{fig:g_cf}. 
We can assume that the optimal outcome of combining the full 10-year data lies in between these two bounds     
$\Neff=184$ and $\Neff=368$.  

\section{Correlating shear measurements across exposures}
\label{sec:cross_correlation}

In addition to the simple averaging scheme discussed in \Sref{sec:Combine}, previous papers \citep[see \eg][]
{2006JCAP...02..001J} also have suggested correlating galaxies in different exposures to eliminate the stochastic 
systematics in the atmosphere and instrument. In theory, by correlating shear measurements across exposures, only the  
non-stochastic systematics remain, which do not scale down further with number of exposures. This, however, 
comes with the price of decreasing the statistical power of a dataset, since there will be a smaller number of pairs that 
contribute to the correlation function. The statistical errors are increased by $\sqrt{\Neff/(\Neff-1)}$.

For LSST, the full 10-year dataset includes $\Neff=$184 -- 368, which means that implementing cross correlation 
will not degrade the statistical power significantly. However, at earlier stages of the survey, when $\Neff$ is still small, 
the $\sqrt{\Neff/(\Neff-1)}$ penalty may overcome the benefit of implementing cross correlation. For the purpose of this 
paper, we will not discuss the details of optimising the combined multi-epoch dataset; rather, we show one simple 
example for implementation of cross correlation and merely demonstrate an alternative way of combining multiple 
exposures to improve upon our results. 

\begin{figure}  
   \begin{center}
   \includegraphics[height=2.4in]{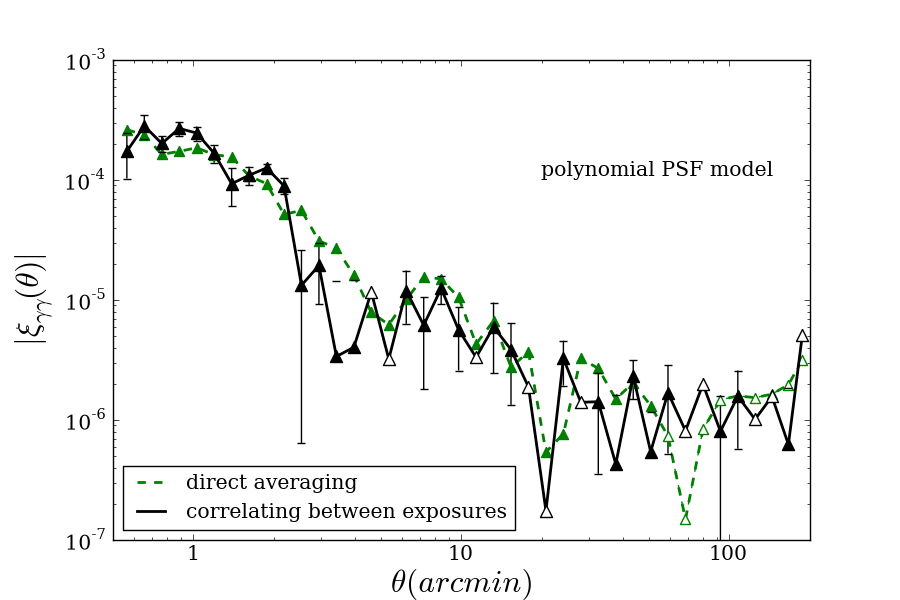}
   \caption{For two typical LSST exposures combined, the spurious shear correlation function when a standard KSB 
   algorithm is implemented with 5th-order polynomial PSF models and only galaxies from different exposures are 
   correlated (black). Compared with the simple average scheme (dashed green), correlating galaxies from different 
   exposures suppresses mainly spurious shear correlation on large scales.  The error bars for the black curve 
   indicate the standard deviation in the 10 curves divided by $\sqrt{10}$. Negative values are plotted with open 
   symbols.}
   \label{fig:cross_cf}  
   \end{center}
 \end{figure}

For the 20 simulations in the master set in \Sref{sec:ShearSys}, we take the spurious shear obtained from the 
5th-order polynomial PSF models and calculate 10 correlation functions by correlating shear 
measurements for galaxies in two different exposures (the 20 exposures were split into 10 groups of image pairs, 
for which the shear measurement in each image is correlated only with shear measurements in the other image). The 
median of the 10 correlation functions is plotted in \Fref{fig:cross_cf}, together with the shear correlation functions 
in \Fref{fig:g_cf}, scaled to $\Neff=2$, so that all curves represent the spurious shear correlation function for combining 
a dataset of two independent exposures. Note that in \Fref{fig:cross_cf}, the shear correlation function from correlating 
galaxies in different exposures is essentially consistent with 
zero at angular scales larger than $\sim3'$, but rises steeply smaller angular scales. This may be because the PSF 
models are consistently ill behaved on the edges of the CCD sensors from the polynomial PSF model, making some of the 
PSF model errors ``non-stochastic'' between frames. This is likely to be an unrealistic artifact since no rotation/dithering is 
used in the 20 images. 


This suggests that correlating galaxies across different exposures can suppress some of the stochastic 
systematics. But similar to the auto-correlation technique, the calculation depends on a good PSF modelling process 
that does not create artificial non-stochastic errors. Given the loss of statistical power by correlating across exposures, 
a more detailed investigation of these tradeoffs will be needed to establish the optimal use of LSST's multi-epoch 
dataset.   

\end{document}